\documentclass[dvipsnames]{lmcs} 
\pdfoutput=1

\usepackage{lastpage}
\lmcsdoi{21}{1}{24}
\lmcsheading{}{\pageref{LastPage}}{}{}%
{Mar.~13,~2024}{Mar.~12,~2025}{}

\keywords{Games under delayed control, delay games, synthesis}

\usepackage{hyperref}
\usepackage[utf8]{inputenc}
\usepackage[T1]{fontenc}
\usepackage{amssymb,amsmath,amsthm}
\usepackage{xspace}
\usepackage{booktabs}

\usepackage[english]{babel}
\addto\extrasenglish{

}

\usepackage{tikz}

\usetikzlibrary{shadows.blur}
\usetikzlibrary{decorations.pathreplacing}
\usetikzlibrary{automata}
\tikzset{>=stealth, shorten >=1pt}
\tikzset{every edge/.style = {thick, ->, draw}}
\tikzset{every loop/.style = {thick, ->, draw}}

\newcommand{\myquot}[1]{``#1''}

\tikzset{mystate/.style={draw,inner sep=3,circle}}

\newcommand{\nats}{\mathbb{N}}
\newcommand{\size}[1]{|#1|}

\renewcommand{\epsilon}{\varepsilon}
\renewcommand{\phi}{\varphi}

\newcommand{\set}[1]{\{#1\}}

\renewcommand{\complement}[1]{\overline{#1}}

\newcommand{\bigo}{\mathcal{O}}

\newcommand{\delaygame}[1]{\Gamma\!_{k}(#1)}
\newcommand{\delaygamep}[1]{\Gamma\!_{k'}(#1)}

\newcommand{\SigmaI}{\Sigma_I}
\newcommand{\SigmaO}{\Sigma_O}

\newcommand{\stratO}{\tau_O}
\newcommand{\stratI}{\tau_I}

\newcommand{\p}{P}

\newcommand{\arenagame}{\mathcal{G}}
\newcommand{\win}{\mathrm{Win}}

\newcommand{\trans}{\rightarrow}
\newcommand{\translabel}[1]{\xrightarrow{{#1}}}

\newcommand{\stratC}{\tau_c}
\newcommand{\stratE}{\tau_e}

\newcommand{\play}{\mathrm{play}}

\newcommand{\pref}[2]{#1[#2]}
\newcommand{\prefs}{\mathrm{Pref}(\arenagame)}
\newcommand{\prefsE}{\mathrm{Pref}_e(\arenagame)}
\newcommand{\prefsC}{\mathrm{Pref}_c(\arenagame)}

\renewcommand{\prob}[1]{{\mathcal{P}}\left(#1\right)}

\newcommand{\aut}{\mathcal{A}}
\newcommand{\init}{I}
\newcommand{\col}{\Omega}

\newcommand{\pspace}{{\upshape{\textsc{PSpace}}}\xspace}
\newcommand{\exptime}{{\upshape{\textsc{ExpTime}}}\xspace}

\newcommand{\twoexp}{{\upshape{\textsc{2ExpTime}}}\xspace}
\newcommand{\threeexp}{{\upshape{\textsc{3ExpTime}}}\xspace}

\newcommand\abbrv[1]{}

\theoremstyle{plain}
\newtheorem{transformation}[thm]{Transformation}

\begin{document}

\title[On the Existence of Reactive Strategies Resilient to Delay]{On the Existence of Reactive Strategies\texorpdfstring{\\}{} Resilient to Delay}

\titlecomment{{\lsuper*}Full version of \cite{FWZ23}, contains all proofs omitted in the conference version as well as a new section on winning games under delayed control with mixed strategies with respect to a fixed threshold~(\autoref{ Sec:Probability})}

\author[M.~Fränzle]{Martin Fränzle\lmcsorcid{0000-0002-9138-8340}}[a]
\author[P.~Kröger]{Paul Kröger\lmcsorcid{0000-0002-0301-3611}}[a]
\author[S.~Winter]{Sarah Winter\lmcsorcid{0000-0002-3499-1995}}[b]
\author[M.~Zimmermann]{Martin Zimmermann\lmcsorcid{0000-0002-8038-2453}}[c]

\address{Carl von Ossietzky Universität\\ Oldenburg, Germany}	
\email{martin.fraenzle@uol.de, p.kroeger@uol.de}  

\address{IRIF \& Université Paris Cité, Paris, France}	
\email{sarah.winter@irif.fr}  

\address{Aalborg University, Aalborg, Denmark}	
\email{mzi@cs.aau.dk}  




\begin{abstract}
  \noindent We compare games under delayed control and delay games, two types of infinite games modelling asynchronicity in reactive synthesis.
  In games under delayed control both players suffer from partial informedness due to symmetrically delayed communication, while in delay games, the protagonist has to grant lookahead to the alter player.
  
  Our first main result, the interreducibility of the existence of sure winning strategies for the protagonist, allows to transfer known complexity results and bounds on the delay from delay games to games under delayed control, for which no such results had been known.  We furthermore analyse existence of randomized strategies that win almost surely, where this correspondence between the two types of games breaks down. In this setting, some games surely won by the alter player in delay games can now be won almost surely by the protagonist in the corresponding game under delayed control, showing that it indeed makes a difference whether the protagonist has to grant lookahead or both players suffer from partial informedness. These results get even more pronounced when we finally address the quantitative goal of winning with a probability in $[0,1]$. We show that for any rational threshold $\theta \in [0,1]$ there is a game that can be won by the protagonist with exactly probability $\theta$ under delayed control, while being surely won by alter in the delay game setting. All these findings refine our original result that games under delayed control are not determined.
\end{abstract}

\maketitle

\section{Introduction}

Two-player zero-sum games of infinite duration are a standard model for the synthesis of reactive controllers, i.e., correct-by-construction controllers that satisfy their specification even in the presence of a malicious environment. 
In such games, the interaction between the controller and the environment is captured by the rules of the game and the specification on the controller induces the winning condition of the game. 
Then, computing a correct controller boils down to computing a winning strategy. 

Often, it is convenient to express the rules in terms of a graph capturing the state-space such that moves correspond to transitions between these states.
The interaction between the controller and the environment then corresponds to a path through the graph and the winning condition is a language of such paths, containing those that correspond to interactions that satisfy the specification on the controller.

In other settings, it is more convenient to consider a slightly more abstract setting without game graphs, so-called Gale-Stewart games~\cite{GS}.
In such games, the players alternatingly pick a sequence of letters, thereby constructing an infinite word.
The winning condition is a language over infinite words, containing the winning words for one player. 
To capture the synthesis problem, the winning condition has to encode both the specification on the controller as well as the rules of interaction. 
It is straightforward to transform a graph-based game into a Gale-Stewart game and a Gale-Stewart game into a graph-based game such that the existence of winning strategies for both players is preserved.

In the most basic setting of synthesis, both the controller and the environment are fully informed about the current state of the game (complete information). 
However, this scenario is not always realistic. 
Thus, much effort has been poured into studying games under incomplete information where the players are only partially informed about the current state of the game.
Here, we are concerned with a special type of partial information designed to capture delays in perception and action. Such delays either render the most recent moves of the opponent invisible to a player or induce a time lag between the selection and the implementation of an own move, respectively.

As a motivating example, consider the domain of cooperative driving: Here, the exchange of information between cars is limited (and therefore delayed) by communication protocols that have to manage the available bandwidth to transfer information between cars. Other delaying factors include, e.g., complex signal processing chains based on computer vision to detect the locations of obstacles.
Thus, decisions have to be made based on incomplete information, which only arrives after some delay.

\subsection{Games under Delayed Control}
Chen et al.~\cite{ChenFLMZ21} introduced (graph) games under delayed control to capture this type of incomplete information. 
Intuitively, assume the players so far have constructed a finite path~$v_0 \cdots v_n$ through the graph. 
Then, the controller has to base their decision on a visible proper prefix $v_0 \cdots v_{n-\delta}$, where $\delta$ is the amount of delay. 
Hence, the suffix~$v_{n-\delta+1} \cdots v_n$ is not yet available to base the decision on, although the decision to be made is to be applied at the last state $v_n$ in the sequence.

Chen et al.\ showed that solving games under delayed control with safety conditions and with respect to  a given delay is decidable: 
They presented two algorithms, an exponential one based on a reduction to delay-free safety games using a queue of length~$\delta$, and a more practical incremental algorithm synthesizing a series of
controllers handling increasing delays and reducing game-graph size in between. 
They showed that even a naïve implementation of this algorithm outperforms the reduction-based one, even when the latter is used with state-of-the-art solvers for delay-free games.
However, the exact complexity of the incremental algorithm and that of solving games under delayed control remained open.

Note that asking whether there is some delay~$\delta$ that allows controller to win reduces to solving standard, i.e., delay-free games, as they correspond to the case~$\delta = 0$. The reason is monotonicity in the delay: if the controller can win for delay $\delta$ then also for any $\delta' < \delta$.
More interesting is the question whether controller wins with respect to  every possible delay. 
Chen et al.\ conjectured that there is some exponential $\delta$ such that if the controller wins under delay~$\delta$, then also under every $\delta'$. 
We call a $\delta$ with this property decisive.

\subsection{Delay Games}
There is also a variant of Gale-Stewart games modelling delayed interaction between the players \cite{DBLP:conf/icalp/HoschL72}. 
Here, the player representing the environment (often called Player~$I$ for input) has to provide a lookahead on their moves, i.e., the player representing the controller (accordingly called Player~$O$ for output) has access to the first $n+k$ letters picked by Player~$I$ when picking their $n$-th letter.
So, $k$ is the amount of lookahead that Player~$I$ has to grant Player~$O$. 
Note that the lookahead benefits Player~$O$ (representing the controller) while the delay in a game under delayed control disadvantages the controller.

Only three years after the seminal Büchi-Landweber theorem showing that delay-free games with $\omega$-regular winning conditions are decidable~\cite{BL69}, Hosch and Landweber showed that it is decidable whether there is a $k$ such that Player~$O$ wins a given Gale-Stewart game with lookahead~$k$~\cite{DBLP:conf/icalp/HoschL72}. 
Forty years later, Holtmann, Kaiser, and Thomas~\cite{HoltmannKT12} revisited these games (and dubbed them delay games). 
They proved that if Player~$O$ wins a delay game then Player~$O$ wins it already with at most doubly-exponential lookahead (in the size of a given deterministic parity automaton recognizing the winning condition).
Thus, unbounded lookahead does not offer any advantage over doubly-exponential lookahead in games with $\omega$-regular winning conditions.
Furthermore, they presented an algorithm with doubly-exponential running time solving delay games with $\omega$-regular winning conditions, i.e., determining whether there exists a $k$ such that Player~$O$ wins a given delay game (with its winning condition again given by a deterministic parity automaton) with lookahead~$k$.

Both upper bounds were improved and matching lower bounds were later proven by Klein and Zimmermann~\cite{KleinZ14}: Solving delay games is \exptime-complete and exponential lookahead is both necessary to win some games and sufficient to win all games that can be won.
Both lower bounds already hold for winning conditions specified by deterministic safety automata while the upper bounds hold for deterministic parity automata.
The special case of solving delay games with conditions given as reachability automata is \pspace-complete, but exponential lookahead is still necessary and sufficient.
Thus, there are tight complexity results for delay games, unlike for games under delayed control.

\subsection{On Player Names}
Note that games under delayed control and delay games are in a sense asymmetric: In a game under delayed control, one models situations in which the controller is disadvantaged by the delayed access to environment's moves while in a delay game, one models situations in which the controller (Player~$O$) benefits from a lookahead on environment's (Player~$I$) moves. 
Said differently, in a delay game, Player~$I$ is disadvantaged by having to provide a lookahead on their moves to Player~$O$. 
Thus, it is natural to ask whether there is a connection between the two disadvantaged players (i.e., controller in a game under delayed control and Player~$I$ in a delay game) and between the other two players (environment and Player~$O$). 

For the sake of consistency with existing literature, we prefer to stick to the player names \myquot{controller} and \myquot{environment} as well as \myquot{Player~$I$} and \myquot{Player~$O$}, even though in a delay game Player~$O$ is typically understood to represent the controller.

Also, let us remark that winning conditions in both types of games are always formulated from the perspective of controller, i.e., for controller in games under delayed control and for Player~$O$ in delay games.
Hence, as we will relate controller and Player~$I$, we always have to complement winning conditions.

\subsection{Our Contributions}

In this work, we prove that there is indeed a tight relation between controller in a game under delayed control and Player~$I$ in a delay game.
More precisely, we show that one can transform a safety game under delayed control in polynomial time into a delay game with a reachability condition for Player~$O$ (i.e., with a safety condition for Player~$I$) such that controller wins the game under delayed control with even delay~$\delta$ if and only if Player~$I$ wins the resulting delay game with lookahead of size~$\frac{\delta}{2}$. 
Dually, we show that one can transform a delay game with safety condition for Player~$I$ in polynomial time into a reachability game under delayed control such that Player~$I$ wins the delay game with lookahead of size~$\delta$ if and only if controller wins the resulting game under delayed control with delay~$2\delta$. 
Thus, we can transfer both upper and lower bound results on complexity and on (necessary and sufficient) lookahead from delay games to delayed control. 
In particular, determining whether controller wins a given safety game under delayed control for every possible delay is \pspace-complete. 
Our reductions also prove the conjecture by Chen et al.\ on the delays that allow controller to win such games: There is an exponential decisive delay in games under delayed control with safety conditions.

Furthermore, we generalize our translation from games with safety conditions to games with reachability conditions, games with parity conditions, and games with winning conditions given by formulas of Linear Temporal Logic (LTL)~\cite{Pnueli}, again allowing us to transfer known results for delay games to games under delayed control.
\autoref{table:results} lists our results.

\begin{table}
    \centering
    \begin{tabular}{lll}
        \toprule
       winning conditon  & complexity  & decisive delay \\
         \midrule
       safety  & \pspace-complete & exponential\\
       reachability  & \exptime-complete & exponential\\
       parity  & \exptime-complete & exponential \\
       LTL  & \threeexp-complete & triply-exponential \\
         \bottomrule
    \end{tabular}
    \caption{Our results for games under delayed control. \myquot{Complexity} refers to the problem of determining whether controller wins the game for every possible delay. All bounds for the decisive delays are tight.}
    \label{table:results}
\end{table}

Note that we have only claimed that the existence of winning strategies for the controller in the game under delayed control and Player~$I$ in the delay game coincides. 
This is no accident! 
In fact, the analogous result for relating environment and Player~$O$ fails. 
This follows immediately from the fact that delay games are determined while games under delayed control are undetermined, even with safety conditions. The reason is that the latter games are truly incomplete information games (which are typically undetermined) while delay games are complete information games. 

We furthermore refine these findings by a detailed comparison between environment and Player~$O$ both in the setting of pure (deterministic) and in the setting of mixed (randomized) strategies. 
The latter setting increases power for both the controller and the environment, making them win (almost surely) games under delayed control that remain undetermined in the deterministic setting, but it also breaks the correspondence between controller and Player~$I$ observed in the deterministic setting: there are games that controller wins almost surely while the corresponding Player~$I$ surely looses them --- and thus Player~$O$, generally corresponding to the environment, wins.

We finally provide results concerning existence of strategies that do not have to win surely or almost surely, but guarantee a win with a given probability $\theta \in ]0,1[$ at least. Again, this relaxation of the winning condition to at least $\theta$-sure winning does not change anything for delay games, as they are determined, meaning that one of the two parties Player~$I$ and Player~$O$ has a sure (and thus also an almost sure and an at least $\theta$-sure) winning strategy. For games under delayed control, the relaxation however makes a tremendous difference: we show that for every rational~$\theta \in [0,1]$ there exists an $\omega$-regular game that the controller, taking its optimal strategy, wins with probability $\theta$ exactly. By symmetry, the same applies to the environment. 

These findings confirm our original intuition that delay games and games under delayed control are indeed different, with the latter in relation giving  more power to controller by symmetrically subjecting both parties, i.e., the environment also, to incomplete information. In the asymmetric setting of delay games, delay comes as a penalty for Player~$I$ only, while it is a benefit for Player~$O$, who is granted a lookahead.

\section{Preliminaries}
\label{sec:prels}

We denote the non-negative integers by~$\nats$.
An {alphabet}~$\Sigma$ is a non-empty finite set of {letters}.
A {word} over $\Sigma$ is a finite or infinite sequence of letters of $\Sigma$: 
The set of finite words (non-empty finite words, infinite words) over $\Sigma$ is denoted by $\Sigma^*$ ($\Sigma^+$, $\Sigma^\omega$).
The {empty word} is denoted by $\varepsilon$,
the length of a finite word~$w$ is denoted by~$\size{w}$.
Given two infinite words~$\alpha \in (\Sigma_0)^\omega$ and $\beta \in (\Sigma_1)^\omega$, we define $\binom{\alpha}{\beta} = \binom{\alpha(0)}{\beta(0)}\binom{\alpha(1)}{\beta(1)}\binom{\alpha(2)}{\beta(2)} \cdots \in (\Sigma_0 \times \Sigma_1)^\omega$. 

\subsection{Games under Delayed Control}
\label{subsec:games}

Games under delayed control are played between two players, controller and environment. 
For pronomial convenience~\cite{McNaughton00}, we refer to controller as she and environment as he.

A game~$\arenagame = ( S, s_0, S_c, S_e, \Sigma_c, \Sigma_e, \trans, \win)$ consists of a finite set~$S$ of states partitioned into the states~$S_c \subseteq S$ of the controller and the states~$S_e \subseteq S$ of the environment, an initial state~$s_0 \in S_c$, the sets of actions~$\Sigma_c$ for the controller and $\Sigma_e$ for the environment, and a transition function~$\trans\colon (S_c \times \Sigma_c) \cup (S_e \times \Sigma_e) \rightarrow S$ such that $s \in S_c$,  $\sigma \in \Sigma_c$ implies $\trans\!\!(s,\sigma) \in S_e$, and $s \in S_e$, $\sigma \in \Sigma_e$ implies $\trans\!\!(s,\sigma) \in S_c$. Finally, $\win \subseteq S^\omega$ is the winning condition of the game.
We write $s \translabel{\sigma} s'$ as shorthand for $s' =\, \trans\!\!(s, \sigma)$.

A play in $\arenagame$ is an infinite sequence~$\pi = \pi_0 \sigma_0  \pi_1 \sigma_1  \pi_2 \sigma_2 \cdots $ satisfying $\pi_0 = s_0$ and $\pi_n \translabel{\sigma_n} \pi_{n+1}$ for all $n \ge 0$.
We say that controller wins $\pi$ if $\pi_0 \pi_1 \pi_2 \cdots \in \win$; otherwise, we say that environment wins $\pi$.
The play prefix of $\pi$ of length~$n$ is defined as $\pref{\pi}{n} = \pi_0 \sigma_0 \cdots \sigma_{n-1} \pi_n$, i.e., $n$ is the number of actions (equivalently, the number of transitions).
We denote by $\prefs$ the set of play prefixes of all plays in $\arenagame$, which is partitioned into the sets~$\prefsC$ and $\prefsE$ of play prefixes ending in $S_c$ and $S_e$, respectively. 
Due to our alternation assumption, play prefixes of even (odd) length are in $\prefsC$ ($\prefsE$).

Fix some even $\delta \ge 0$. A strategy for the controller in $\arenagame$ under delay~$\delta$ is a pair $(\alpha, \stratC)$ where $\alpha \in (\Sigma_c)^{\frac{\delta}{2}}$ and $\stratC \colon \prefsC \rightarrow \Sigma_c$ maps play prefixes ending in $S_c$ to actions of the controller.
A play~$\pi_0 \sigma_0 \pi_1 \sigma_1 \pi_2 \sigma_2 \cdots$ is consistent with $(\alpha,\stratC)$ if $\sigma_0 \sigma_2 \cdots \sigma_{\delta -4} \sigma_{\delta-2} = \alpha$ and $\sigma_{2n} = \stratC(\pref{\pi}{2n - \delta})$ for all $2n > \delta -2$, i.e., controller has access to environment's actions with a delay of $\delta$.
In particular, her first $\frac{\delta}{2} +1$ actions are independent of environment's actions and, in general, her $n$-th action~$\sigma_{2n}$ only depends on the actions~$\sigma_1, \sigma_3, \ldots, \sigma_{(2n-\delta) -1}$ picked by environment, but not on the actions~$\sigma_{(2n-\delta)+1},\sigma_{(2n-\delta)+3},\ldots, \sigma_{2n-1}$ picked by environment.
The strategy~$(\alpha, \stratC)$ is winning under delay~$\delta$ if every play that is consistent with it is winning for controller.
Controller wins $\arenagame$ under delay~$\delta$ if she has a winning strategy under delay~$\delta$ for $\arenagame$.

\begin{rem}
\label{remark:delaycontrol} \hfill
 \begin{enumerate}
    \item 
The notion of winning strategy for controller under delay~$0$ is the classical one for delay-free games (cf.~\cite{GTW02}).
    \item
\label{remarkitem:delaycontrol:monotonicity}  If 
controller wins $\arenagame$ under delay~$\delta$, then also under every delay~$\delta' < \delta$~\cite{ChenFLMZ21}.
 \end{enumerate}
\end{rem}
\noindent 
A strategy for environment is a mapping~$\stratE\colon \prefsE \rightarrow \Sigma_e$.
A play~$\pi_0 \sigma_0 \pi_1 \sigma_1 \pi_2 \sigma_2 \cdots$ is consistent with $\stratE$ if $\sigma_{2n+1} = \stratE(\pi_0 \sigma_0 \cdots \sigma_{2n-1} \pi_{2n+1})$ for all $n \ge 0$, i.e., environment has access to the full play prefix when picking his next action.
The strategy~$\stratE$ is winning, if every play that is consistent with it is winning for the environment (i.e., the sequence of states is not in $\win$).
Further, we say that environment wins $\arenagame$, if he has a winning strategy for $\arenagame$. 
Note that the two definitions of strategies are in general not dual, e.g., the one for environment is not defined with respect to a delay~$\delta$. 
 \begin{rem}
 The notion of winning strategy for environment is the classical one for delay-free games (cf.~\cite{GTW02}).
 \end{rem}

We say that a game under delayed control~$\arenagame$ is determined under delay~$\delta$, if either controller wins $\arenagame$ under delay~$\delta$ or environment wins $\arenagame$.
Let us stress that determinacy is defined with respect to some fixed~$\delta$ and that $\arenagame$ may be determined for some $\delta$, but undetermined for some other~$\delta'$ (due to the non-dual definition of strategies).
\autoref{remark:undetermined} shows an undetermined safety game under delayed control.

\begin{exa}
\label{example:gameunderdelayedcontrol}
Consider the game~$\arenagame = (S, s_0, S_c, S_e, \Sigma_c, \Sigma_e, \rightarrow, \win)$ depicted in \autoref{fig:gameunderdelayedcontrol} where $\win $ contains all plays that do not visit the black vertex.
Note that this is a safety condition. 
In particular, if controller does not pick action~$b$ at $c_2$ and does not pick action~$a$ at $c_3$, then the vertex $e_3$ is never reached.
This is straightforward without delay, but we claim that controller can also win $\arenagame$ under delay~$2$.

\begin{figure}
    \centering
\begin{tikzpicture}[thick,scale=1.3]
    
    \draw[gray!20, rounded corners,line width=2mm] (-1.15,-1.8) rectangle (10.35,2.2);  
    
    \draw[fill=gray!20,gray!20] (1.5,-1.8) rectangle (4.5,2.2); 
    \draw[fill=gray!20,gray!20] (7.5,-1.8) rectangle (10.3,2.2); 
    
    \node[black] at (0,1.9) {\small \bf C};
    \node[black] at (3,1.9) {\small \bf E};
    \node[black] at (6,1.9) {\small \bf C};
    \node[black] at (9,1.9) {\small \bf E};

    \node[mystate] (c1) at (0,0) {$c_1$};
    \node[mystate] (e1) at (3,1) {$e_1$};
    \node[mystate] (e2) at (3,-1) {$e_2$};
    \node[mystate] (c2) at (6,1) {$c_2$};
    \node[mystate] (c3) at (6, -1) {$c_3$};
    \node[mystate] (e4) at (9,1) {$e_4$};
    \node[mystate,fill=black,text=white] (e3) at (3,0) {$e_3$};
    \node[mystate] (e5) at (9,-1) {$e_5$};

     \path[-stealth]
     (-.75,0) edge (c1)
     (c1) edge[near start] node[above] {$a$} (e1)
     (c1) edge[near start] node[below] {$b$} (e2)
     (e1) edge[near start] node[below] {$u,u'$} (c2)
     (e2) edge[near start] node[above] {$u,u'$} (c3)
     (c2) edge[near start] node[below] {$a$} (e4)
     (c2) edge[near start] node[below] {$b$} (e3)
     (c3) edge[near start] node[above] {$a$} (e3)
     (c3) edge[near start] node[above] {$b$} (e5)
     (e3) edge[] node[above,near start] {$u,u'$} (c1)
     (e4) edge[bend right =21] node[above,pos=.07] {$u$} (c1.north)
     (e4) edge[bend right=0] node[very near start,below] {$u'$} (c3)
     (e5) edge[bend left =21] node[above,below,pos=.07] {$u$} (c1.south)
     (e5) edge[bend left=0] node[above,very near start] {$u'$} (c2)
    ;

\end{tikzpicture}
    \caption{The game for \autoref{example:gameunderdelayedcontrol}. Controller wins all plays that never visit the black vertex. Note that we have $\Sigma_c =\set{a,b} $ and $\Sigma_e = \set{u,u'}$.}
    \label{fig:gameunderdelayedcontrol}
\end{figure}
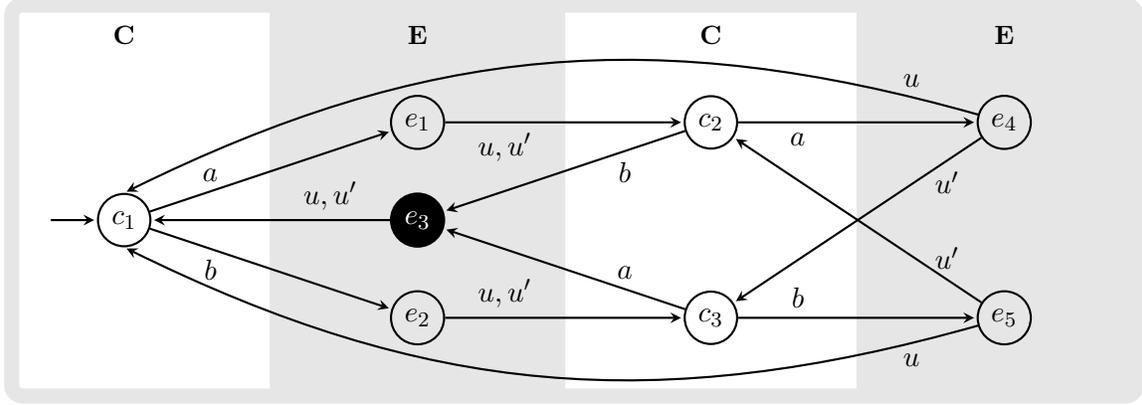

To gain some intuition, consider a play prefix~$\pi_0 \sigma_0 \pi_1 \cdots \pi_{n-1}\sigma_{n-1}\pi_{n}$ with $n\ge 4$ and $\pi_n \in S_c$.
Then, controller has to pick an action~$\sigma_n$ to continue the prefix.
However, due to the delayed control, she has to do so based on the prefix~$\pi_0 \sigma_0 \pi_1 \cdots \pi_{n-3}\sigma_{n-3}\pi_{n-2}$.

If $\pi_{n-2}$ is $c_2$, then $\pi_n$ is either $c_3$ or $c_1$. 
Hence, picking $\sigma_n = b$ is the only safe choice.
Dually, if $\pi_{n-2}$ is $c_3$, then $\pi_n$ is either $c_2$ or $c_1$. 
Hence, picking $\sigma_n = a$ is the only safe choice.

Finally, assume $\pi_{n-2}$ is $c_1$.
Then, $\pi_n$ is either $c_2$ or $c_3$. 
In the former case, picking $\sigma_n = a$ is the only safe choice, in the latter case, picking $\sigma_n = b$ is the only safe choice. 
So, controller needs to distinguish these two cases, although she has no access to $\pi_n$.

But she can do so by inspecting~$\pi_{n-3}$ (which she has access to): 
As a predecessor of $\pi_{n-2} = c_1$, it can either be $e_4$, $e_5$, or $e_3$.
In the latter case, the play is already losing.
Thus, we disregard this case, as we construct a winning strategy. 
So, assume we have $\pi_{n-3} = e_4$ (the case~$\pi_{n-3} = e_5$ is dual).
Then, we must have $\pi_{n-4} = c_2$ (the only predecessor of $e_4$) and, by our analysis of the safe moves above, controller must have picked $\sigma_{n-2} = b$ (based, due to delay, on the prefix ending in $\sigma_{n-4} = c_2$). 
From this we can conclude $\pi_{n-1} = e_2$ and thus $\pi_n = c_3$ (the only successor of $e_2$).
Thus, she can safely pick $\sigma_{n} = b$.

This intuition, and the necessary initialization, is implemented by the strategy~$(\alpha,\stratC)$ with $\alpha = a$ and 
\[
\stratC(\pi_0 \sigma_0 \pi_1 \cdots \pi_{n-3}\sigma_{n-3}\pi_{n-2}) = \begin{cases}
a &\text{$n = 2$ and $\pi_0 = c_1$,} \\
b&\text{$n > 2$, $\pi_{n-2} = c_1$, and $\pi_{n-3} = e_4$,} \\
a&\text{$n > 2$, $\pi_{n-2} = c_1$, and $\pi_{n-3} = e_5$,} \\
b&\text{$\pi_{n-2} = c_2$,} \\
a&\text{$\pi_{n-2} = c_3$}.
\end{cases}
\]
An induction over the play length shows that $(\alpha,\stratC)$ is winning for controller under delay~$2$.
\end{exa}

\begin{rem}
Our definition of games under delayed control differs in three aspects from the original definition of Chen et al.~\cite{ChenFLMZ21}.
\begin{itemize}

    \item We allow arbitrary winning conditions while Chen et al.\ focused on safety conditions.
    
    \item The original definition allows nondeterministic strategies (a strategy that returns a nonempty set of actions, each one of which can be taken), while we restrict ourselves here to deterministic strategies (a strategy that returns a single action to be taken). 
    The motivation for their use of nondeterministic strategies is the fact that they can be refined if additional constraints are imposed, which Chen et al.'s algorithm computing a winning strategy relies on.
    
    Here, on the other hand, we are just interested in the existence of winning strategies.
    In this context, it is sufficient to consider deterministic strategies, as controller has a nondeterministic winning strategy if and only if she has a deterministic winning strategy.
    Also, strategies in delay games are deterministic, so the transformation between games under delayed control and delay games can be formulated more naturally for deterministic strategies.
    
    \item The original definition also allowed odd delays~$\delta$ while we only allow even delays. 
    As we will see in \autoref{sec:transformation}, the transformation of games under delayed control to delay games is naturally formulated for even delays.
    This choice also simplifies definitions, as accounting for odd delays imposes an additional notational burden.
    
\end{itemize}
\end{rem}

\subsection{Delay Games}
\label{subsec:delaygames}

{Delay games} are played between two players, Player~$I$ (she) and Player~$O$ (he). 
A delay game~$\delaygame{L}$ (with constant lookahead) consists of a lookahead~$k \in \nats$ and a {winning condition}~$L \subseteq (\SigmaI \times \SigmaO)^\omega$ for some alphabets~$\SigmaI$ and $\SigmaO$. 
Such a game is played in rounds~$n = 0,1,2, \ldots$ as follows: in round~$0$, first Player~$I$ picks a word~$x_0 \in \SigmaI^{k+1}$, then Player~$O$ picks a letter~$y_0 \in \SigmaO$. 
In round~$n>0$, Player~$I$ picks a letter~$x_n \in \SigmaI$, then Player~$O$ picks a letter~$y_n \in \SigmaO$.
So, a play of $\delaygame{L}$ has the form~$(x_0, y_0)(x_1, y_1)(x_2, y_2) \cdots $, where $x_0$ is a word of length~$k+1$ and all other $x_n$ (those with $n >0$) and all $y_n$ are letters.
Player~$O$ {wins} such a play~$(x_0, y_0)(x_1, y_1)(x_2, y_2) \cdots $ if the outcome~$\binom{ x_0 x_1 x_2 \cdots }{ y_0 y_1 y_2 \cdots }$ is in $L$; otherwise, Player~$I$ wins.

A {strategy} for Player~$I$ in $\delaygame{L}$ is a mapping $\stratI \colon \SigmaO^* \rightarrow \SigmaI^*$ satisfying $\size{\stratI(\epsilon)} = k+1$ and $\size{\stratI(w)} = 1$ for all $w \in \SigmaO^+$. 
A strategy for Player~$O$ is a mapping~$\stratO \colon \SigmaI^+ \rightarrow \SigmaO$. A play~$(x_0, y_0)(x_1, y_1)(x_2, y_2) \cdots $ is {consistent} with $\stratI$ if $x_n = \stratI(y_0 \cdots y_{n-1})$ for all $n \ge 0$, and it is consistent with $\stratO$ if $y_n = \stratO(x_0 \cdots x_n)$ for all $n \ge 0$. 
So, strategies are dual in delay games, i.e., Player~$I$ has to grant some lookahead on her moves that Player~$O$ has access to.
A strategy for Player~$\p \in \set{I,O}$ is {winning}, if every play that is consistent with the strategy is won by Player~$\p$.
We say that Player~$\p\in \set{I,O}$ {wins} a game  $\delaygame{L}$ if Player~$\p$ has a winning strategy in $\delaygame{L}$.

\begin{rem}
\label{remark:delaymonotonicity}
 \hfill
 \begin{itemize}
     \item 
    If Player~$O$ wins $\delaygame{L}$, then he also wins $\delaygamep{L}$ for every $k' > k$.
     \item 
    If Player~$I$ wins $\delaygame{L}$, then she also wins $\delaygamep{L}$ for every $k' < k$.
 \end{itemize}
\end{rem}
\noindent 
Unlike games under delayed control, delay games with Borel winning conditions are determined~\cite{KleinZ14}, i.e., each delay game~$\delaygame{L}$ with Borel~$L$ and fixed~$k$ is won by one of the players.

\begin{exa}
\label{example:delaygame}
Consider 
\[
L=\left\{ \binom{a_0}{b_0}\binom{a_1}{b_1}\binom{a_2}{b_2}\cdots \mid b_0 \notin\set{a_0, a_1, a_2}  \right\}\] over the alphabets~$\SigmaI = \SigmaO = \set{1,2,3,4}$.

Player~$I$ wins $\delaygame{L}$ for $k=1$ with the following strategy~$\stratI$: $\stratI(\epsilon) = 12$ and $\stratI(b_0) = b_0$, and $\stratI(w)$ arbitrary for all $w \in \SigmaO^+$ with $\size{w} > 1$: In round~$0$, after Player~$I$ has picked~$a_0a_1 = 12$, Player~$O$ has to pick some $b_0$. In order to not loose immediately, he has to pick $b_0 \notin \set{1,2}$. Then, in round~$1$, Player~$I$ picks $a_2 = b_0$ and thereby ensures $b_0 \in \set{a_0, a_1, a_2}$. 
Hence, the play is not won by Player~$O$ (it's outcome is not in $L$), therefore it is winning for Player~$I$.

However, Player~$O$ wins $\delaygame{L}$ for $k = 2$ with the following strategy~$\stratO$: $\stratO(a_0a_1a_2)$ is a letter in the nonempty set~$\SigmaO \setminus \set{a_0, a_1, a_2}$ and $\stratO(w)$ arbitrary for all $w \in \SigmaI^+$ with $\size{w} \neq 3$.
In round~$0$, after Player~$I$ has picked~$a_0a_1a_2$, Player~$O$ picks $b_0 \notin \set{a_0, a_1, a_2}$ and thus ensures that the outcome is in $L$.
\end{exa}

\begin{rem}
We restrict ourselves here to the setting of constant lookahead, i.e., in a delay game~$\delaygame{L}$ in round~$n$ when Player~$O$ picks her $n$-th letter, Player~$I$ has already picked $k+n+1$ letters (note that we start in round~$0$ with the zeroth letter).
Delay games have also been studied with respect to growing lookahead, i.e., the lookahead increases during a play~\cite{HoltmannKT12}.
However, it is known that constant lookahead is sufficient for all $\omega$-regular winning conditions: if Player~$O$ wins for any lookahead (no matter how fast it is growing), then she also wins with respect to constant lookahead, which can even be bounded exponentially in the size of a deterministic parity automaton recognizing the winning condition~\cite{KleinZ14}.
Stated differently, growing lookahead does not allow to win any more games than constant lookahead.
Finally, the setting of constant lookahead in delay games considered here is the natural counterpart to games under delayed control, where the delay is fixed during a play. 
\end{rem}

\subsection{Games under Delayed Control vs.\ Delay Games}
\label{subsec:comparison}

Let us illustrate the differences between the strategic abilities of the players in a game~$\arenagame$ under delayed control (say with delay~$\delta$) and the players in a delay game~$\delaygame{L}$. 

Let us start with the delay game~$\delaygame{L}$. 
Here, Player~$I$ has to grant a lookahead of $k$ moves to Player~$O$, who can base her moves on this additional information. 
Formally, consider a play prefix~$(x_0, y_0)\cdots (x_n,y_n)$, keeping in mind that $x_0$ is a word of length~$k+1$ and each $x_{n'}$ with $n' >0$ and each $y_{n'}$ is a single letter. 
Hence, Player~$I$ has picked $x_0 x_1 \cdots x_n$ of length~$n+1+k$ while Player~$O$ has picked $y_0 y_1 \cdots y_n$ of length~$n+1$.
Now, Player~$I$'s next move~$x_{n+1}$ may depend on $y_0 y_1 \cdots y_n$ and Player~$O$'s subsequent move depends on $x_0 x_1 \cdots x_n x_{n+1}$.
Thus, a delay game is a game of complete information, both players have full access to the moves made by their respective opponent.\footnote{As usual, a player can always reconstruct their own previous moves, if necessary.}

On the other hand, in the game~$\arenagame$ under delayed control, controller has to base her moves on a proper prefix of the play prefix constructed thus far while environment does not gain a symmetric advantage from controller's disadvantage.
To understand this, consider a play prefix~$\pi_0 \sigma_0 \pi_1 \cdots \pi_{n-1} \sigma_{n-1} \pi_n$ with (w.l.o.g.) $n > \delta$.
If $\pi_n \in S_c$, i.e., it is controllers turn, then she has to base her decision on the prefix~$\pi_0 \sigma_0 \pi_1 \cdots \pi_{n-1-\delta} \sigma_{n-1-\delta} \pi_{n- \delta}$, picking some action~$\sigma_{n}$ leading to a state~$\pi_{n+1} \in S_e$.
Thus, the actions~$\sigma_{n+1-\delta}, \sigma_{n+3-\delta},\ldots, \sigma_{n-1}$ have already been picked by environment, but are not available to controller to base her decision on.
Thus, she has incomplete information about the evolution of the play.
Crucially, environment does not benefit from controller's disadvantage:
Environment has to pick action~$\sigma_{n}$ only based on the actions~$\sigma_0,\ldots, \sigma_{n-1}$ while controller has to pick the action~$\sigma_{n+\delta}$ based on the actions~$\sigma_0,\ldots, \sigma_{n-1}$.

We will study this difference in informedness of the players in a game under delayed control and in a delay game in the setting of deterministic and randomized strategies.

\subsection{\texorpdfstring{$\mathbf{\omega}$-Automata}{omega-Automata}}
\label{subsec:automata}

A deterministic reachability automaton~$\aut = (Q, \Sigma, q_\init, \delta_\aut, F)$ consists of a finite set~$Q$ of states containing the initial state~$q_\init \in Q$ and the set of accepting states~$F \subseteq Q$, an alphabet~$\Sigma$, and a transition function~$\delta_\aut \colon Q \times \Sigma \rightarrow Q$.
The size of $\aut$ is defined as $\size{\aut} = \size{Q}$.
Let $w = w_0 w_1 w_2 \cdots \in \Sigma^\omega$.
The run of $\aut$ on $w$ is the sequence~$q_0 q_1 q_2 \cdots $ such that $q_0 = q_\init$ and $q_{n+1} = \delta_\aut(q_n, w_n)$ for all $n\ge 0$.
A run~$q_0q_1 q_2\cdots$ is (reachability) accepting if $q_n \in F$ for some $n \ge 0$.
The language (reachability) recognized by $\aut$, denoted by $L(\aut)$, is the set of infinite words over $\Sigma$ such that the run of $\aut$ on $w$ is (reachability) accepting.

A deterministic safety automaton has the form~$\aut = (Q, \Sigma, q_\init, \delta_\aut, U)$ where $Q, \Sigma, q_\init, \delta_\aut$ are as in a deterministic reachability automaton and where $U \subseteq Q$ is a set of unsafe states.
The notions of size and runs are defined as for reachability automata, too. 
A run~$q_0q_1q_2 \cdots$ is (safety) accepting if $q_n \notin U$ for all $n \ge 0$.
The language (safety) recognized by $\aut$, again denoted by $L(\aut)$, is the set of infinite words over $\Sigma$ such that the run of $\aut$ on $w$ is (safety) accepting.

Reachability and safety automata accept each a fragment of the $\omega$-regular languages, while parity automata accept exactly the $\omega$-regular languages~\cite{GTW02}. 
A deterministic parity automaton has the form~$\aut = (Q, \Sigma, q_\init, \delta_\aut, \col)$ where $Q, \Sigma, q_\init, \delta_\aut$ are as in a deterministic reachability automaton and where $\col \colon Q \rightarrow \nats$ is a coloring of the states.
The notions of size and runs are defined as for reachability automata, too. 
A run~$q_0q_1q_2 \cdots$ is (parity) accepting if the maximal color appearing infinitely often in the sequence~$\col(q_0)\col(q_1)\col(q_2)\cdots $ is even.
The language (parity) recognized by $\aut$, again denoted by $L(\aut)$, is the set of infinite words over $\Sigma$ such that the run of $\aut$ on $w$ is (parity) accepting.

Reachability and safety automata are dual while parity automata are self-dual.

\begin{rem}
\label{remark:automataduality}
 \hfill
 \begin{enumerate}
     \item 
    Let $\aut = (Q, \Sigma, q_\init,\delta_\aut, F)$ be a deterministic reachability automaton and let $\complement{\aut}$ be the deterministic safety automaton~$(Q, \Sigma, q_\init,\delta_\aut, F)$. Then, $L(\complement{\aut}) = \complement{L(\aut)}$.
    
     \item 
    Let $\aut = (Q, \Sigma, q_\init,\delta_\aut, F)$ be a deterministic safety automaton and let $\complement{\aut}$ be the deterministic reachability automaton~$(Q, \Sigma, q_\init,\delta_\aut, F)$. Then, $L(\complement{\aut}) = \complement{L(\aut)}$.
    
     \item 
    Let $\aut = (Q, \Sigma, q_\init,\delta_\aut, \col)$ be a deterministic parity automaton and let $\complement{\aut}$ be the deterministic parity automaton~$(Q, \Sigma, q_\init,\delta_\aut, q \mapsto \col(q)+1)$. Then, $L(\complement{\aut}) = \complement{L(\aut)}$.
 \end{enumerate}
\end{rem}

\section{From Games under Delayed Control to Delay Games and Back}
\label{sec:transformation}

In this section, we exhibit a tight correspondence between controller in games under delayed control and Player~$I$ in delay games. 
Recall that in a game under delayed control, it is the controller whose control is delayed, i.e., she is at a disadvantage as she only gets delayed access to the action picked by environment.
In a delay game, it is Player~$I$ who is at a disadvantage as she has to grant a lookahead on her moves to Player~$O$. 
Thus, when simulating a game under delayed control by a delay game, it is natural to let Player~$I$ take the role of controller and let Player~$O$ take the role of environment. 
Also recall that the winning condition~$\win$ in a game under delayed control is formulated from controller's point-of-view: the winning condition requires her to enforce a play in $\win$. 
On the other hand, the winning condition~$L$ of a delay game is formulated from the point-of-view of Player~$O$: Player~$O$ has to enforce a play whose outcome is in $L$. 
Thus, as Player~$I$ takes the role of controller, we need to {complement} the winning condition to reflect this change in perspective: 
The set of winning outcomes for Player~$I$ in the simulating delay game is the complement of $\win$.

In the remainder of this section, we show how to simulate a game under delayed control by a delay game and then the converse, i.e., we show how to simulate a delay game by a game under delayed control.

\begin{transformation}
\label{trans:controltodelaygame}
First, we transform a game under delayed control into a winning condition for a delay game, i.e., a language. 
In a delay game with this winning condition, the players simulate a play in the game under delayed control by picking actions, which uniquely induce such a play.
To formalize this, we need to introduce some notation.
Fix a game~$\arenagame = (S, s_0, S_c, S_e, \Sigma_c, \Sigma_e, \trans,\win)$.
Note that a sequence~$\sigma_0 \sigma_1 \sigma_2 \cdots \in (\Sigma_c \Sigma_e)^\omega$ induces a unique play~$\play(\sigma_0 \sigma_1 \sigma_2 \cdots ) = \pi_0 \sigma_0 \pi_1 \sigma_1 \pi_2 \sigma_2 \cdots$ in $\arenagame$ which is defined as follows:
$\pi_0 = s_0$ and $\pi_{n+1} = \trans\!\!(\pi_n, \sigma_n)$ for all $n \ge 0$.
Likewise, a finite sequence~$\sigma_0 \sigma_1 \cdots \sigma_n \in (\Sigma_c \Sigma_e)^*(\Sigma_c + \varepsilon)$ induces a unique play prefix~$\play(\sigma_0 \sigma_1 \cdots \sigma_n)$ which is defined analogously.

Now, we define the language~$L(\arenagame) \subseteq (\Sigma_c \times \Sigma_e)^\omega$ such that $ \binom{\sigma_0}{\sigma_1} \binom{\sigma_2}{\sigma_3} \binom{\sigma_4}{\sigma_5} \cdots \in L(\arenagame)$  if and only if $\play(\sigma_0 \sigma_1 \sigma_2 \cdots )$ is winning for controller.
\end{transformation}

\begin{exa}
We continue \autoref{example:gameunderdelayedcontrol} to illustrate the transformation: consider the game under delayed control~$\arenagame = (S, s_0, S_c, S_e, \Sigma_c, \Sigma_e, \rightarrow, \win)$ depicted in \autoref{fig:gameunderdelayedcontrol} on Page~\pageref{example:gameunderdelayedcontrol} where $\win $ contains all plays that do not visit the black vertex. 
The language~$L(\arenagame)$ is accepted by the safety automaton depicted in \autoref{fig:reachautforexample}, which accepts an infinite sequence of actions (arranged in tuples) if and only if the unique induced play in $\arenagame$ is winning for controller.
In a delay game with winning condition~$\overline{L(\arenagame)}$ (recall we need to complement the winning condition), Player~$I$ picks actions from $\Sigma_c$, Player~$O$ picks actions from $\Sigma_e$, and Player~$I$ wins if and only if the induced play in $\arenagame$ is in $\win$. 

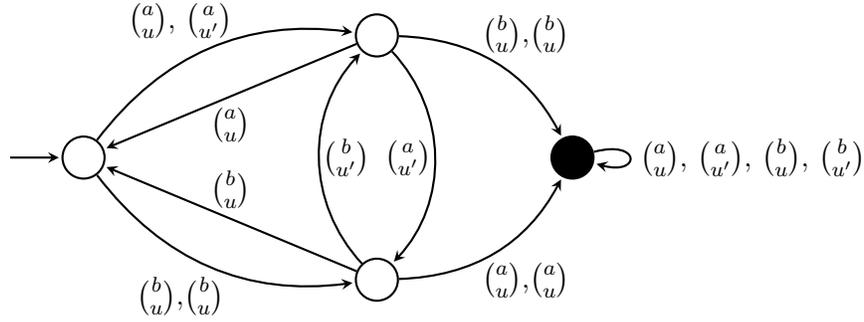
\begin{figure}
    \centering
\begin{tikzpicture}[thick,scale=1.3]
        
    \node[mystate] (c1) at (3,0) {\phantom{v}};
    \node[mystate] (c2) at (6,1.25) {\phantom{v}};
    \node[mystate] (c3) at (6, -1.25) {\phantom{v}};
    \node[mystate,fill=black,text=white] (e3) at (8,0) {\phantom{v}};

     \path[-stealth]
     (2.25,0) edge (c1)
     (c1) edge[bend left] node[above,xshift=-.4cm] {$\binom{a}{u}$, $\binom{a}{u'}$} (c2)
     
     (c1) edge[bend right] node[below,xshift=-.4cm] {$\binom{b}{u}$,$\binom{b}{u'}$} (c3)
 
     (c2) edge[bend left=0] node[below] {$\binom{a}{u}$} (c1)
          
     (c2) edge[bend left=43] node[left,xshift=.1cm] {$\binom{a}{u'}$} (c3)

     (c2) edge[bend left] node[above,xshift=.4cm] {$\binom{b}{u}$,$\binom{b}{u'}$} (e3)
     
     (c3) edge[bend right] node[below, xshift=.4cm] {$\binom{a}{u}$,$\binom{a}{u'}$} (e3)
     
     (c3) edge[bend left=0] node[above] {$\binom{b}{u}$} (c1)
     
     (c3) edge[bend left=43] node[right,xshift=-.1cm] {$\binom{b}{u'}$} (c2)

     (e3) edge[loop right] node[right] {$\binom{a}{u}$, $\binom{a}{u'}$, $\binom{b}{u}$, $\binom{b}{u'}$ } ()
     
     ;

\end{tikzpicture}
    \caption{The safety automaton accepting the winning condition~$L(\arenagame)$ obtained by applying \autoref{trans:controltodelaygame} to the game from \autoref{example:gameunderdelayedcontrol}. A run is accepting if it does not visit the black state.}
    \label{fig:reachautforexample}
\end{figure}
\end{exa}

Now, we prove the correspondence between $\arenagame$ and $\delaygame{\complement{L(\arenagame)}}$. 
The winning condition of the delay game is the complement of $L(\arenagame)$, which implements the switch of perspective described above. 

\begin{lem}\label{lemma:fromDelayedControlToDelay}
Let $\arenagame$ be a game and $\delta \ge 0$ even. Controller wins $\arenagame$ under delay $\delta$ if and only if Player~$I$ wins $\delaygame{\complement{L(\arenagame)}}$ for $k = \frac{\delta}{2}$.
\end{lem}

\begin{proof}
To simplify our notation in the following, we denote $\delaygame{\complement{L(\arenagame)}}$ by $\Gamma$.

First, let $(\alpha,\stratC)$ be a winning strategy for controller under delay $\delta$ in $\arenagame$. 
We inductively define a strategy~$\stratI$ for Player~$I$ in $\Gamma$.

For the induction start, we define $\stratI(\varepsilon) = \alpha\cdot \stratC(s_0)$, where $s_0$ is the initial vertex of $\arenagame$. Note that $|\alpha| = k$, i.e., $\stratI(\varepsilon)$ has the required length.
For the induction step, we define $\stratI(\sigma_1\sigma_3\cdots\sigma_{2n+1}) = \stratC(\play(\sigma_0\sigma_1 \cdots\sigma_{2n+1}))$ for all $n \ge 0$, where $\sigma_0\sigma_2\cdots\sigma_{\delta-2} = \alpha$ and $\sigma_{2i} = \stratI(\sigma_1\sigma_3\cdots\sigma_{2i-1})$ for all $2i$ in the range $\delta-2 \le 2i \le 2n$.

We show that $\stratI$ is a winning strategy for Player~$I$ in $\Gamma$.
Consider a play in $\Gamma$ that is consistent with $\stratI$, say with outcome $\binom{ \sigma_0 \sigma_2 \sigma_4 \cdots }{ \sigma_1 \sigma_3 \sigma_5 \cdots }$.
We show that the play is winning for Player~$I$ by showing that its outcome is in $L(\arenagame)$, i.e., $\play(\sigma_0\sigma_1\sigma_2\cdots)$ is winning for controller in $\arenagame$.
To this end, consider the unique play $\pi$ in $\arenagame$ under delay $\delta$ that is consistent with $\stratC$ and where environment plays such that his moves spell $\sigma_1\sigma_3\cdots$.
By definition of $\stratI$, we have $\pi = \play(\sigma_0\sigma_1\sigma_2\cdots)$.
Since $\stratC$ is a winning strategy for controller, we indeed obtain that $\pi$ is winning for controller in $\arenagame$.

For the other direction, let $\stratI$ be a winning strategy for Player~$I$ in $\Gamma$.
We define a strategy $(\alpha,\stratC)$ for controller in $\arenagame$ under delay $\delta$.
Let $\stratI(\varepsilon) = a_0\cdots a_k$.
We define $\alpha = a_0\cdots a_{k-1}$ and $\stratC(\pi_0) = a_k$. Note that $|\alpha| =  \frac{\delta}{2}$, i.e., $\alpha$ has the required length.
Furthermore, we define 
$\stratC(\pi_0\sigma_0\pi_1\cdots\sigma_{2n+1}\pi_{2n+2}) = \stratI(\sigma_1\sigma_3\cdots\sigma_{2n+1})$ for all play prefixes~$\pi_0\sigma_0\pi_1\cdots\sigma_{2n+1}\pi_{2n+2}$ of length~$2n+2$ for $n \ge 0$.

We show that $(\alpha,\stratC)$ is a winning strategy for controller in $\arenagame$ under delay $\delta$.
Consider a play~$\pi = \pi_0\sigma_0\pi_1\sigma_1\pi_2\sigma_2\cdots$ consistent with $(\alpha,\stratC)$.
We show that $\pi$ is winning for controller in $\arenagame$.  
To this end, consider the unique play in $\Gamma$ that is consistent with $\stratI$ and where Player~$O$ plays such that his moves spell $\sigma_1\sigma_3\cdots$.
By definition of $\stratC$, the outcome of this play is $\binom{ \sigma_0 \sigma_2 \sigma_4 \cdots }{ \sigma_1 \sigma_3 \sigma_5 \cdots }$.
Since $\stratI$ is a winning strategy for Player~$I$ in $\Gamma$, we obtain $\binom{ \sigma_0 \sigma_2 \sigma_4 \cdots }{ \sigma_1 \sigma_3 \sigma_5 \cdots } \notin \complement{L(\arenagame)}$, i.e., \par\vspace*{0.25\baselineskip}\noindent$\binom{ \sigma_0 \sigma_2 \sigma_4 \cdots }{ \sigma_1 \sigma_3 \sigma_5 \cdots } \in L(\arenagame)$. 
Hence, $\play(\sigma_0\sigma_1\sigma_2\cdots) = \pi$ is indeed winning for controller in $\arenagame$.
\end{proof}

\noindent 
Now, we consider the converse and transform a delay game into a game under delayed control.

\begin{transformation}
\label{trans:delaygametocontrol}
Fix a delay game~$\delaygame{L}$ with $L \subseteq (\SigmaI \times \SigmaO)^\omega$. 
We construct a game under delayed control to simulate $\delaygame{L}$ as follows: The actions of controller are the letters in $\SigmaI$, and the actions of environment are the letters in $\SigmaO$. Thus, by picking actions, controller and environment construct the outcome of a play of $\delaygame{L}$.
As winning conditions of games under delayed control only refer to states visited by a play, but not the actions picked by the players, we reflect the action picked by a player in the state reached by picking that action.
Here, we have to require without loss of generality that $\SigmaI$ and $\SigmaO$ are disjoint.

Formally, we define~$\arenagame(L) = (S, s_I, S_c,S_e,\Sigma_c, \Sigma_e, \trans,\win)$ with $S = S_c \cup S_e$, $S_c = \set{s_I} \cup \SigmaO$, $S_e = \SigmaI$, $\Sigma_c = \SigmaI$, $\Sigma_e = \SigmaO$, $\trans\!\!(s, a) = a$ for all $s \in S_c$ and $a \in \SigmaI$, and $\trans\!\!(s,b) =  b$ for all $s \in S_e$ and $b \in \SigmaO$.
Finally, we define $\win = \set{s_I s_0 s_1 s_2\cdots \mid \binom{s_0}{s_1}\binom{s_2}{s_3}\binom{s_4}{s_5}\cdots \in L}$.
\end{transformation}

\begin{exa}
We continue \autoref{example:delaygame} to illustrate the second transformation:
consider the delay game~$\delaygame{L}$ with
\[L=\left\{ \binom{a_0}{b_0}\binom{a_1}{b_1}\binom{a_2}{b_2}\cdots \mid b_0 \notin\set{a_0, a_1, a_2}  \right\}\] over the alphabets~$\SigmaI = \SigmaO = \set{1,2,3,4}$.
Note that this game does not satisfy our assumption that $\SigmaI$ and $\SigmaO$ are disjoint. 
Hence, in the following we decorate all letters in $\SigmaO$ by a prime and consider the winning condition \[L = \left\{ \binom{a_0}{b_0}\binom{a_1}{b_1}\binom{a_2}{b_2}\cdots \mid b_0 \notin\set{a_0', a_1', a_2'}  \right\}\] over the alphabets~$\SigmaI = \set{1,2,3,4}$ and $\SigmaO = \set{1',2',3',4'}$.

Now, applying \autoref{trans:delaygametocontrol} yields the game under delayed control depicted in \autoref{fig:controlgameforexample}.
In this game, controller picks letters from $\SigmaI$ (by moving to the corresponding state) and environment picks letters from $\SigmaO$ (again by moving to the corresponding state). 
Thus, controller and environment construct the outcome of a play of $\delaygame{L}$.
Now, the winning condition of $\arenagame(L)$ is defined such that controller wins $\arenagame(\overline{L})$ (recall that we need to complement the winning condition) if and only if the resulting outcome is in $L$.

\begin{figure}
    \centering
\begin{tikzpicture}[thick,scale=1.3]
    
    \draw[gray!20, rounded corners,line width=2mm] (-1.15,0.35) rectangle (7.5,5.2);  
    
    \draw[fill=gray!20,gray!20] (1.5,0.35) rectangle (4.5,5.2); 
    
    \node[black] at (0,4.7) {\small \bf C};
    \node[black] at (3,4.7) {\small \bf E};
    \node[black] at (6,4.7) {\small \bf C};

    \node[mystate] (s0) at (0,2.5) {$s_I$};
    
    \foreach \i in {1,2,3,4} {
    \node[mystate] (e\i) at (3,-\i+5) {$\i$};
    }

    \foreach \i in {1,2,3,4} {
    \node[mystate] (c\i) at (6,-\i+5) {$\i'$};
    }

     \path[-stealth]
     (-.7,2.5) edge (s0)
    ; 
    \foreach \i in {1,2,3,4} {
     \path[-stealth]
     (s0) edge node[above] {} (e\i);
    }

    \foreach \i in {1,2,3,4} {
    \foreach \j in {1,2,3,4} {

     \path[-stealth]
     (e\i) edge (c\j)
     (c\i) edge (e\j);

    }}

    ;

\end{tikzpicture}
    \caption{The game under delayed control obtained by applying \autoref{trans:delaygametocontrol} to the delay game from \autoref{example:delaygame}. All incoming edges to vertex~$j$ are labeled by action~$j$, all incoming edges to vertex~$j'$ are labeled by action~$j'$.}
    \label{fig:controlgameforexample}
\end{figure}
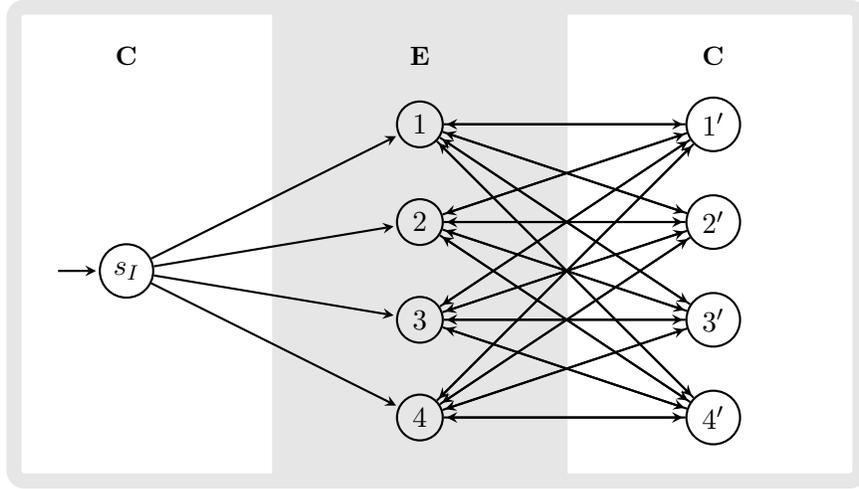

\end{exa}

The following remark simplifies the proof of correctness of the second transformation.
It follows by a careful inspection of the definitions.

\begin{rem}
\label{remark:transformationsareinverses}
Let $L \subseteq (\SigmaI \times \SigmaO)^\omega$. Then, $L = {L(\arenagame(L))}$.
\end{rem}

Now, we show that the second transformation is correct, again using complementation to implement the perspective switch.

\begin{lem}
\label{lemma:fromDelayToDelayedControl}
Let $L \subseteq (\SigmaI\times\SigmaO)^\omega$ and $k \ge 0$. Player~$I$ wins $\delaygame{L}$ if and only if controller wins $\arenagame(\complement{L})$ under delay~$2k$.
\end{lem}

\begin{proof}
Consider the delay game $\delaygame{\complement{L(\arenagame(L)}}$.
By \autoref{lemma:fromDelayedControlToDelay}, controller wins $\arenagame(\complement{L})$ under delay $2k$ if and only if Player~$I$ wins $\delaygame{\complement{L(\arenagame(\complement{L})}}$.
Applying \autoref{remark:transformationsareinverses} yields $\complement{L} = {L(\arenagame(\complement{L}))}$ and thus ${L} = \complement{L(\arenagame(\complement{L}))}$.
Hence, as $\delaygame{L}$ and $\delaygame{\complement{L(\arenagame(\complement{L})}}$ have the same winning condition, Player~$I$ wins the former if and only if she wins the latter.
Thus, the statement of the lemma directly follows.
\end{proof}

\section{Results}
\label{sec:results}

\autoref{lemma:fromDelayedControlToDelay} and \autoref{lemma:fromDelayToDelayedControl} allow us to transfer results from delay games to games under delayed control.
Due to the definitions of strategies in games under delayed control not being dual, we first consider controller and then environment.

Recall that delay that allows controller to win satisfies a monotonicity property (see \autoref{remark:delaycontrol}(\ref{remarkitem:delaycontrol:monotonicity})): if controller wins a game under delay~$\delta$, then also under every delay~$\delta' < \delta$.
Thus, the set of delays for which controller wins is downward-closed, i.e., it is either a finite set~$\set{0,2,4,\ldots, \delta_{\max}}$ or it is equal to the set~$2\nats$ of even numbers.
In the following, we study the complexity of determining whether controller wins under all possible delays, whether she wins under a given delay, and determine bounds on~$\delta_{\max}$.

Note that winning for environment is independent of delay and boils down to the classical notion of winning delay-free games~\cite{GTW02}, which is a well-studied problem.
Hence, we disregard this problem.
However, we do discuss the relation between environment in a game under delayed control and Player~$O$ in the simulating delay game constructed in the previous section.


Before we present our results, we need to specify how to measure the size of games and delay games, especially how winning conditions are represented (recall that, so far, they are just $\omega$-languages).
In the following, we only consider $\omega$-regular winning conditions specified by $\omega$-automata (see \autoref{subsec:automata}) or formulas of Linear Temporal Logic (LTL)~\cite{Pnueli}, which subsume the typical specification languages for winning conditions.
Hence, the size of a game~$(S, s_0, S_c, S_e, \Sigma_c, \Sigma_e, \trans, \win)$ under delayed control is given by the sum~$\size{S} + \size{\Sigma_c} + \size{\Sigma_e} + \size{\win}$, where $\size{\win}$ is the size of an automaton or LTL formula (measured in the number of distinct subformulas) representing $\win$, depending on whether $\win$ is specified by an automaton or a formula.
Analogously, for a delay game~$\delaygame{L}$, we define the size of $L$ as the size of an automaton or LTL formula (measured in the number of distinct subformulas) representing $L$.
The bound~$k$ is encoded in binary, if necessary.

\subsection{Safety}

A game~$\arenagame = (S, s_0, S_c, S_e, \Sigma_c, \Sigma_e, \trans, \win)$ with winning condition~$\win$ is a safety game if $\win$ is accepted by a deterministic safety automaton. 

\begin{rem}
When Chen et al.\ introduced safety games under delayed control, they did not use automata to specify their winning plays, but instead equipped the game with a set of unsafe states and declared all those plays winning for controller that never visit an unsafe state.
It is straightforward to see that our definition is equivalent, as their definition is captured by a deterministic safety automaton with two states.
Conversely, taking the product of a game and a deterministic safety automaton yields an equivalent game with a state-based safety condition. 
\end{rem}
\noindent 
Our results rely on the following two bounds on the transformations presented in \autoref{sec:transformation}, which are obtained by applying \autoref{remark:automataduality}:
\begin{enumerate}
    \item If the winning condition~$\win$ for a game~$\arenagame$ under delayed control is given by a deterministic safety automaton with $n$ states, then the winning condition~$\complement{L(\arenagame)}$ is recognized by a deterministic reachability automaton with $n$ states.
    
    \item Dually, if the winning condition~$L \subseteq (\SigmaI \times \SigmaO)^\omega$ of a delay game is given by a deterministic reachability automaton with $n$ states, then the winning condition of the game~$\arenagame(\complement{L})$ under delayed control is recognized  by a deterministic safety automaton with $\bigo(n\cdot\size{\SigmaI})$ states.
    \label{page:safety}Here, the factor~$\size{\SigmaI}$ stems from the fact that an automaton~$\aut$ for $L$ processes pairs~$(a,b)\in \SigmaI \times \SigmaO$ of letters while an automaton for the winning condition of $\arenagame(\complement{L})$ needs to simulate $\aut$ while first receiving $a$ and then $b$, i.e., it needs to store $a \in \SigmaI$ in its state. 
\end{enumerate}
\noindent 
We begin by settling the complexity of determining whether controller wins a given safety game under every delay, which follows from the \pspace-completeness of determining whether there is a lookahead that allows Player~$O$ to win a given delay game with reachability winning condition~\cite{KleinZ14}.

\begin{thm}
\label{thm:controlleruniversality}
The following problem is \pspace-complete: Given a safety game~$\arenagame$, does controller win $\arenagame$ under every delay~$\delta$?
\end{thm}

\begin{proof}
The following problem is \pspace-complete~\cite{KleinZ14}: Given a deterministic reachability automaton $\aut$, does Player~$O$ win $\delaygame{L(\aut)}$ for some $k$?
Hence, due to determinacy of delay games with $\omega$-regular winning conditions \cite{KleinZ15} and due to closure of \pspace under complementation, the following problem is also \pspace-complete: Given a deterministic reachability automaton $\aut$, does Player~$I$ win $\delaygame{L(\aut)}$ for all $k$?
Thus, the \pspace lower bound follows from \autoref{lemma:fromDelayToDelayedControl} and the corresponding upper bound from \autoref{lemma:fromDelayedControlToDelay}.
\end{proof}

\noindent 
Next, we give a lower bound on the complexity of determining whether controller wins a given safety game under a given delay, which is derived from a lower bound for delay games with reachability winning conditions.

\begin{thm}
\label{thm:fixeddelta}
The following problem is \pspace-hard: Given a safety game~$\arenagame$ and $\delta$ (encoded in binary), does controller win $\arenagame$ under delay~$\delta$.
\end{thm}

\begin{proof}
The following problem is \pspace-hard \cite{KleinZ14}\footnote{The result is not stated as such, but follows by inspecting the proof of Theorem~4.1 in \cite{KleinZ14}, which shows that Player~$O$ wins a delay game with reachability condition with respect to any lookahead if and only if he wins it with respect to constant lookahead~$2^{\size{\aut}}$.}: Given a deterministic reachability automaton $\aut$, does Player~$O$ win $\delaygame{L(\aut)}$ for $k = 2^{\size{\aut}}$?
Hence, due to determinacy of delay games with $\omega$-regular winning conditions \cite{KleinZ15} and due to closure of \pspace under complementation, the following problem is also \pspace-complete: Given a deterministic reachability automaton $\aut$, does Player~$I$ win $\delaygame{L(\aut)}$ for $k = 2^{\size{\aut}}$?
Thus, the \pspace lower bound follows from \autoref{lemma:fromDelayToDelayedControl} using $\delta = 2k$.
\end{proof}

\noindent 
Note that we do not claim any upper bound on the problem considered in \autoref{thm:fixeddelta}.
There is a trivial \twoexp upper bound obtained by hardcoding the delay into the graph of the safety game, thereby obtaining a classical delay-free safety game. 
It is open whether the complexity can be improved. 
Let us remark though that, via the correspondence to delay games presented in \autoref{sec:transformation}, improvements here would also yield improvements on the analogous problem for delay games, which is open too~\cite{Zimmermann22}.

Next, we turn our attention to bounds on the delay for which controller wins.
Recall that due to monotonicity, the set of delays for which controller wins is downward-closed, i.e., it is either a finite set~$\set{0,2,4,\ldots, \delta_{\max}}$ or it is equal to $2\nats$.
In the following, we present tight bounds on the value~$\delta_{\max}$.

As a consequence, we settle a conjecture by Chen et al.: They conjectured that there is some delay~$\delta_t$ (exponential in $\size{\arenagame}$), such that if controller wins $\arenagame$ under delay~$\delta_t$, then she wins under every delay.
Note that this conjecture implies that $\delta_{\max}$ is at most exponential.

The following theorem proves Chen et al.'s conjecture, while \autoref{thm:delaylb} shows that $\delta_t$ must necessarily be exponential.
For $\delta_{\max}$ this means it is at most exponential for every game, and can be exponential for some games.

The following two results are again obtained from similar bounds for delay games with reachability winning conditions.

\begin{thm}
\label{thm:universaldelta}
Let $\arenagame$ be a safety game. 
There is a $\delta_t \in \bigo(2^{\size{\arenagame}})$ such that if controller wins $\arenagame$ under delay~$\delta_t$, then she wins $\arenagame$ under every $\delta$.
\end{thm}

\begin{proof}
Exponential lookahead is sufficient for Player~$O$ to win delay games with reachability conditions:
Let $\aut$ be a deterministic reachability automaton with $n$ states recognizing a language~$L$. 
If Player~$O$ wins $\delaygame{L}$ for some $k$, then he also wins $\delaygamep{L}$ for $k' = 2^{n}$~\cite{KleinZ14}.
So, dually, if Player~$O$ does not win $\delaygamep{L}$ for $k' = 2^n$, then he does not win $\delaygame{L}$ for any $k$.
Due to determinacy~\cite{KleinZ15}, this is equivalent to the following statement: if Player~$I$ wins $\delaygamep{L}$ for $k' =2^n $, then she wins $\delaygame{L}$ for every $k$.
We transfer this result to safety games under delayed control.

Given $\arenagame$, let $\delta_t = 2\cdot 2^{\size{\arenagame}}$.
Now, assume controller wins $\arenagame$ under delay~$\delta_t$. 
Then, Player~$I$ wins $\delaygame{\complement{L(\arenagame)}}$ for $k = \frac{\delta_t}{2} =  2^{\size{\arenagame}}$ due to \autoref{lemma:fromDelayedControlToDelay}.
As argued above, $\complement{L(\arenagame)}$ is recognized by a deterministic reachability automaton of size at most~$\size{\arenagame}$.
Thus, Player~$I$ wins $\delaygame{\complement{L(\arenagame)}}$ for every $k$. 
Hence, due to \autoref{lemma:fromDelayToDelayedControl}, controller wins $\arenagame$ under every $\delta$.
\end{proof}

\noindent 
Finally, we show that the exponential upper bound on $\delta_{\max}$ is tight.

\begin{thm}
\label{thm:delaylb}
For every $n > 1$, there is a safety game~$\arenagame_n$ of size~$\bigo(n)$ such that  controller wins $\arenagame$ under delay~$2^n$, but not under delay~$2^n+2$.
\end{thm}

\begin{proof}
For every~$n > 1$, there is a deterministic reachability automaton~$\aut_n$ of size~$\bigo(n)$ and where $\SigmaI$ is of size~$n$ such that Player~$O$ wins $\delaygame{L(\aut_n)}$ for $k = 2^n$, but Player~$I$ wins $\delaygamep{L(\aut_n)}$ for every $k' < 2^n$~\cite{KleinZ14}.
Now, we consider the safety games~$\arenagame(L(\aut_n))$, which have size~$\bigo(n)$ as argued above.
Applying \autoref{lemma:fromDelayToDelayedControl} shows that controller wins $\arenagame(L(\aut_{n}))$ under delay~$2\cdot2^n$.
Now, towards a contradiction, assume controller wins $\arenagame(L(\aut_n))$ under delay~$2^n+2$. 
Then, \autoref{lemma:fromDelayedControlToDelay} yields that Player~$I$ wins $\delaygame{{L(\arenagame(L(\aut_n)))}} = \delaygame{L(\aut_n)}$ (see \autoref{remark:transformationsareinverses}) for $k = 2^n$, yielding the desired contradiction.
\end{proof}

\subsection{Reachability}
In this subsection, we consider the case of winning conditions given by reachability automata. 
A game~$\arenagame = (S, s_0, S_c, S_e, \Sigma_c, \Sigma_e, \trans, \win)$ with winning condition~$\win$ is a reachability game if $\win$ is accepted by a deterministic reachability automaton. 

Applying \autoref{remark:automataduality} yields the following two bounds on the transformations from \autoref{sec:transformation}:
\begin{enumerate}
    \item If the winning condition~$\win$ for a game~$\arenagame$ under delayed control is given by a deterministic reachability automaton with $n$ states, then the winning condition~$\complement{L(\arenagame)}$ is recognized by a deterministic safety automaton with $n$ states.
    
    \item Dually, if the winning condition~$L \subseteq (\SigmaI \times \SigmaO)^\omega$ of a delay game is given by a deterministic safety automaton with $n$ states, then the winning condition of the game~$\arenagame(\complement{L})$ under delayed control is recognized  by a deterministic reachability automaton with $\bigo(n\cdot\size{\SigmaI})$ states.
    Here, the additional factor~$\size{\SigmaI}$ is again required to store an input letter, as in the case of safety automata (see Page~\pageref{page:safety}).
\end{enumerate}
Exponential lookahead is both sufficient to win all delay games with safety conditions that can be won and required to win some of these games~\cite{KleinZ14}.
Furthermore, determining whether there is some lookahead that allows Player~$O$ to win a given delay game with safety condition is \exptime-complete~\cite{KleinZ14}.
As in the case of safety games, we can transfer these results to games under delayed control with $\omega$-regular winning conditions.

\begin{thm}
\hfill
\begin{enumerate}
    \item The following problem is \exptime-complete: Given a reachability game~$\arenagame$, does controller win $\arenagame$ under every delay~$\delta$?
    \item Let $\arenagame$ be a reachability game with winning condition specified by a deterministic reachability automaton with $n$ states. 
There is a $\delta_t \in \bigo(2^{n^2})$ such that if controller wins $\arenagame$ under delay~$\delta_t$, then she wins $\arenagame$ under every $\delta$.
    \item For every $n > 1$, there is a reachability game~$\arenagame_n$ of size $\bigo(n^2)$ with a winning condition specified by a two-state deterministic reachability automaton~$\aut_n$ such that controller wins $\arenagame$ under delay~$2^n$, but not under delay~$2^n+2$.
\end{enumerate}
\end{thm}

\subsection{Parity}

Next, we consider the case of $\omega$-regular winning conditions, given by deterministic parity automata.
Applying \autoref{remark:automataduality} yields the following two bounds on the transformations from \autoref{sec:transformation}:
\begin{enumerate}
    \item If the winning condition~$\win$ for a game~$\arenagame$ under delayed control is given by a deterministic parity automaton with $n$ states, then the winning condition~$\complement{L(\arenagame)}$ is recognized by a deterministic parity automaton with $n$ states.
    
    \item Dually, if the winning condition~$L \subseteq (\SigmaI \times \SigmaO)^\omega$ of a delay game is given by a deterministic parity automaton with $n$ states, then the winning condition of the game~$\arenagame(\complement{L})$ under delayed control is recognized  by a deterministic parity automaton with $\bigo(n\cdot\size{\SigmaI})$ states.
    Here, the additional factor~$\size{\SigmaI}$ is again required to store an input letter, as in the case of safety automata (see Page~\pageref{page:safety})
\end{enumerate}
Exponential lookahead is both sufficient to win all $\omega$-regular delay games that can be won and required to win some of these games~\cite{KleinZ14}.
Furthermore, determining whether there is some lookahead that allows Player~$O$ to win a given $\omega$-regular delay game is \exptime-complete~\cite{KleinZ14}.
As in the case of safety games, we can transfer these results to games under delayed control with $\omega$-regular winning conditions.

\begin{thm}
\hfill
\begin{enumerate}
    \item The following problem is \exptime-complete: Given a game~$\arenagame$ with $\omega$-regular winning condition specified by a deterministic parity automaton, does controller win $\arenagame$ under every delay~$\delta$?
    \item Let $\arenagame$ be a game with $\omega$-regular winning condition specified by a deterministic parity automaton with $n$ states. 
There is a $\delta_t \in \bigo(2^{n^3})$ such that if controller wins $\arenagame$ under delay~$\delta_t$, then she wins $\arenagame$ under every $\delta$.

    \item For every $n > 1$, there is a game~$\arenagame_n$ of size $\bigo(n^2)$ with $\omega$-regular winning condition specified by a two-state deterministic parity automaton~$\aut_n$ such that controller wins $\arenagame$ under delay~$2^n$, but not under delay~$2^n+2$.
\end{enumerate}
\end{thm}
\noindent 
 Note that the lower bound on $\delta_t$ is just a restatement of \autoref{thm:delaylb}, as safety games have $\omega$-regular winning conditions.

\subsection{Linear Temporal Logic}

Finally, one can also transfer the triply-exponential upper and lower bounds on the necessary lookahead in delay games with LTL winning conditions as well as the \threeexp-completeness of determining whether Player~$O$ wins such a delay game with respect to some lookahead~\cite{KleinZ16} to games under delayed control with LTL winning conditions.
Here, we exploit the following facts:
\begin{enumerate}
    \item If the winning condition~$\win$ for a game~$\arenagame$ under delayed control is given by an LTL formula~$\varphi$, then the winning condition~$\complement{L(\arenagame)}$ is given by an LTL formula of size~$\bigo(\size{\varphi})$.
    
    \item Dually, if the winning condition~$L \subseteq (\SigmaI \times \SigmaO)^\omega$ of a delay game is given by an LTL formula~$\varphi$, then the winning condition of the game~$\arenagame(\complement{L})$ under given action is given by an LTL formula of size $\bigo(\size{\varphi})$.
\end{enumerate}

\begin{thm}
\hfill
\begin{enumerate}
    \item The following problem is \threeexp-complete: Given a game~$\arenagame$ with winning condition specified by an LTL formula~$\varphi$, does controller win $\arenagame$ under every delay~$\delta$?
    \item Let $\arenagame$ be a game with $\omega$-regular winning condition specified by an LTL formula~$\varphi$. 
There is a $\delta_t \in \bigo(2^{2^{2^{\size{\varphi} + \size{\arenagame}}}})$ such that if controller wins $\arenagame$ under delay~$\delta_t$, then she wins $\arenagame$ under every $\delta$.

    \item For every $n > 1$, there is a game~$\arenagame_n$ of size $\bigo(n^2)$ with winning condition specified by an LTL formula~$\varphi_n$ of size~$\bigo(n^2)$ such that controller wins $\arenagame$ under delay~$2^{2^{2^n}}$, but not under delay~$2^{2^{2^n}}+2$.
\end{enumerate}
\end{thm}

\subsection{Environment's View}
\label{sec:environment}

In \autoref{sec:transformation}, we proved a tight correspondence between controller in a game under delayed control and Player~$I$ in a delay game.
Thus, it is natural to ask whether environment and Player~$O$ also share such a tight correspondence.
A first indication that this is not the case can be obtained by considering the determinacy of these games:
While delay games with Borel winning conditions are determined~\cite{KleinZ15}, even safety games under delayed control are not necessarily determined~\cite{ChenFLMZ21}.

Upon closer inspection, the lack of a correspondence between environment and Player~$O$ is not surprising, as the strategies in games under delayed control are not dual between the players:
controller is at a disadvantage as she only gets delayed access to the actions picked by environment while environment does not benefit from this disadvantage. 
He does not get access to the actions picked by controller in advance.
In a delay game however, the strategy definitions are completely dual: Player~$I$ has to grant lookahead on her moves which Player~$O$ gets access to.
Thus, environment is in a weaker position than Player~$O$.\footnote{The difference can be formalized in terms of the information the players have access to: games under delayed control are incomplete-information games while delay games are complete-information games. Although interesting, we do not pursue this angle any further.}

In this section, we study the correspondence between environment and Player~$O$ in detail by formally proving that environment is weaker than Player~$O$.\footnote{Note that the following lemma can easily be proven using determinacy of delay games and the correspondence between controller and Player~$I$ exhibited by \autoref{lemma:fromDelayedControlToDelay} and \autoref{lemma:fromDelayToDelayedControl}, if the winning condition is Borel. However, we state (and prove) the lemma for arbitrary winning conditions.}

\begin{lem}
\label{lemma:envtoinput}
Let $\arenagame$ be a game. If environment wins $\arenagame$ then Player~$O$ wins $\delaygame{\complement{L(\arenagame)}}$ for every $k$.
\end{lem}

\begin{proof}
Fix some $k \geq 0$.
To simplify our notation in the following, we denote $\delaygame{\complement{L(\arenagame)}}$ by $\Gamma$.

Let $\stratE \colon \prefsE \rightarrow \Sigma_c$ be a winning strategy for environment in $\arenagame$. 
We inductively define a strategy $\stratO \colon \SigmaI^+ \rightarrow \SigmaO$ for Player~$O$ in $\Gamma$, which is a straightforward simulation of $\stratE$ ignoring the granted lookahead of $k$ additional letters.

For the induction start, we define $\stratO(\sigma_0\sigma_2\cdots\sigma_{2k}) = \stratE(\play(\sigma_0))$.
For the induction step, we define $\stratO(\sigma_0\sigma_2\cdots\sigma_{2n}) = \stratE(\play(\sigma_0\sigma_1\cdots\sigma_{2(n-k)}))$ for all $n \ge k$, where $\sigma_{2i+1} = \stratO(\sigma_0\sigma_2\cdots\sigma_{2(i+k)})$ for all $2i$ in the range $0 \le 2i \le 2(n-k)-2$.

We show that $\stratO$ is a winning strategy for Player~$O$ in $\Gamma$.
Consider a play in $\Gamma$ that is consistent with $\stratO$, say with outcome $\binom{ \sigma_0 \sigma_2 \sigma_4 \cdots }{ \sigma_1 \sigma_3 \sigma_5 \cdots }$.
We show that the play is winning for Player~$O$ by showing that its outcome is not in $L(\arenagame)$, i.e., $\play(\sigma_0\sigma_1\sigma_2\cdots)$ is winning for environment in $\arenagame$.
To this end, consider the unique play $\pi$ in $\arenagame$ that is consistent with $\stratE$ and where controller plays such that her moves spell $\sigma_0\sigma_2\cdots$.
By definition of $\stratO$, we have $\pi = \play(\sigma_0\sigma_1\sigma_2\cdots)$.
Since $\stratE$ is a winning strategy for environment, we indeed obtain that $\pi$ is winning for environment in $\arenagame$.
\end{proof}

\noindent 
Now, we show that the converse direction of \autoref{lemma:envtoinput} fails.

\begin{lem}
\label{lemma:envfailure}
There is a safety game~$\arenagame$ such that Player~$O$ wins $\delaygame{\complement{L(\arenagame)}}$ for some $k$, but environment does not win $\arenagame$ under any delay.
\end{lem}

\begin{proof}
Let $\arenagame$ be the safety game depicted in \autoref{fig:env}. 
With each move, the players place a coin (by either picking heads or tails) and the environment wins a play by correctly predicting the second action of controller with his first action.
Note that environment has only one nontrivial choice to make, i.e., the choice of head or tails for the first action picked by him.

Clearly, environment has no winning strategy in $\arenagame$ (under any delay) because he has no access to future moves of controller.
Stated differently, if environment picks $h$ ($t$) in his first move, then the play in which the second action of controller is $t$ ($h$) is winning for controller.\footnote{Note that under any delay~$\delta > 0$, controller cannot do this strategically, as she has to fix her first two actions in advance. But as environment has no access to these fixed actions, he cannot react to them strategically. Also, see \autoref{remark:undetermined}.}

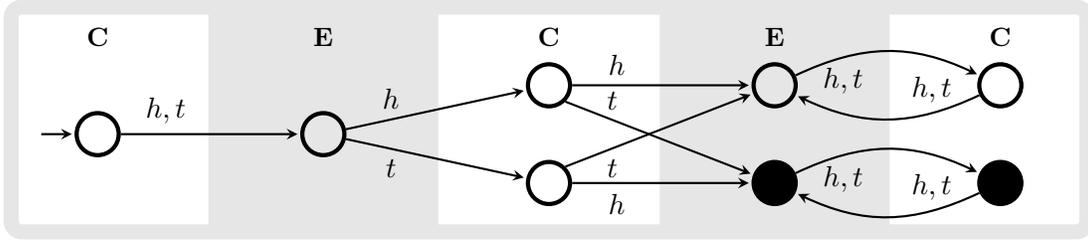
\begin{figure}
    \centering
\begin{tikzpicture}[ultra thick,yscale=1.3]
    
    \draw[gray!20, rounded corners,line width=2mm] (-1.15,-1) rectangle (13.15,1.3);  
    
    \draw[fill=gray!20,gray!20] (1.5,-1) rectangle (4.5,1.3); 
    \draw[fill=gray!20,gray!20] (7.5,-1) rectangle (10.5,1.3); 
    
    \node[black] at (0,1) {\small \bf C};
    \node[black] at (3,1) {\small \bf E};
    \node[black] at (6,1) {\small \bf C};
    \node[black] at (9,1) {\small \bf E};
    \node[black] at (12,1) {\small \bf C};
    
    \node[mystate] (1) at (0,0) {\phantom{v}};
    \node[mystate] (2) at (3,0) {\phantom{v}};
    \node[mystate] (3) at (6,-.5) {\phantom{v}};
    \node[mystate] (4) at (6, .5) {\phantom{v}};
    
    \node[mystate,fill = black] (5) at (9,-.5) {\phantom{v}};
    \node[mystate] (6) at (9, .5) {\phantom{v}};
    
    \node[mystate,fill = black] (7) at (12,-.5) {\phantom{v}};
    \node[mystate] (8) at (12, .5) {\phantom{v}};
    
    \path[-stealth]
    (-.75,0) edge (1)
    (1) edge[near start] node[above] {$h,t$} (2)
    
    (2) edge[near start] node[above] {$h$} (4)
    (2) edge[near start] node[below] {$t$} (3)
    
    (3) edge[near start] node[below] {$h$} (5)  
    (3.north east) edge[near start] node[below] {$t$} (6)  

    (4) edge[near start] node[above] {$h$} (6)  
    (4.south east) edge[near start] node[above] {$t$} (5)  

    (5) edge[bend left=20, near start] node[below] {$h,t$} (7)  
    (7) edge[bend left=20, near start] node[above] {$h,t$} (5)  

    (6) edge[bend left=20, near start] node[below] {$h,t$} (8)  
    (8) edge[bend left=20, near start] node[above] {$h,t$} (6)
    ;

\end{tikzpicture}
    \caption{A safety game that environment does not win under any delay, but Player~$O$ wins the associated delay game with $k \ge 1$. The initial state is marked by an arrow and the unsafe vertices are black. Note that both players have the actions~$h$ and $t$ available.}
    \label{fig:env}
\end{figure}

Now, we consider the delay game~$\delaygame{\complement{L(\arenagame)}}$ for $k = 1$.
Recall that the winning condition~$\complement{L(\arenagame)}$ contains the winning plays for Player~$O$, i.e., we have $\binom{ \sigma_0 \sigma_2 \sigma_4 \cdots }{ \sigma_1 \sigma_3 \sigma_5 \cdots} \in \complement{L(\arenagame)}$ if and only if $\sigma_{1} \neq \sigma_{2}$.
It is easy to see that Player~$O$ has a winning strategy in $\delaygame{\complement{L(\arenagame)}}$ by simply flipping the second letter picked by Player~$I$.
This is possible since Player~$I$ has to provide two letters during the first round.
\end{proof}

\begin{rem}
\label{remark:undetermined}
The safety game~$\arenagame$ depicted in \autoref{fig:env} is in fact undetermined under every delay~$\delta > 0$.
In the proof of \autoref{lemma:envfailure}, we have already established that environment does not win $\arenagame$.
Now, under every delay~$\delta > 0$, controller has to fix at least two actions before getting access to the first action picked by environment. 
This implies that there is, for every strategy for controller under delay~$\delta$, at least one consistent play that is losing for her, i.e., a play in which environment picks $h$ ($t$) if the second move fixed by controller is $t$ ($h$).
Thus, no strategy is winning for controller under delay $\delta$.

Let us remark that, according to our definition of environment strategies, he is not able to enforce a losing play for controller (the game is  undetermined after all), as he does not get access to the second action fixed by controller.
Also, this is again the difference to delay games: Player~$O$ has access to these first two actions when making his first move, and is thereby able to win.
\end{rem}

\begin{figure}
    \centering
    \begin{tikzpicture}[ultra thick,yscale=1.2]

        \begin{scope}
            \draw[fill=gray!10, blur shadow={shadow blur steps=10}](0,-.55)circle[x radius=5cm, y radius=.6cm];
            \clip(0,-.55)circle[x radius=5.1cm, y radius=.619cm];
            \draw[] (-1.6,-1.55) -- (-1.6,.75);
            \draw[] (1.6,.75) -- (1.6,-1.55);
        \end{scope}

        \begin{scope}
                
            \draw[fill=gray!10, blur shadow={shadow blur steps=10}](0,-2.5)circle[x radius=5cm, y radius=.6cm];
            \clip(0,-2.5)circle[x radius=5.1cm, y radius=.619cm];    
            \draw[] (0,-1.5) -- (0,-4.5);
        \end{scope}
        
        \node[] at (0,-.55) {\large undetermined};
        \node[] at (-3,-.55) {\large C wins};
        \node[] at ( 3,-.55) {\large E wins};
        \node[] at (-2, -2.5) {\large $I$ wins};
        \node[] at ( 2, -2.5) {\large $O$ wins};

        \path[-stealth]
        (-3.25,-.75) edge[line width= 1mm,bend right,gray] (-3.25,-2.35)
        (-2.75,-2.25) edge[line width= 1mm,bend right,gray] (-2.75,-.75)
        (3.25,-.75) edge[line width= 1mm,gray] (2.2,-2.3)
        (0,-.95) edge[gray, line width=1mm] (1,-2.3)
        (1.6,-2.2) edge[line width= 1mm,gray,-] (1.6,-1.7) 
        (1.6,-1.8) edge[line width= 1mm,gray] (2.2,-.75)
        (1.6,-1.8) edge[line width= 1mm,gray] (1,-.75)
        ;
        
        \node[align=left,anchor=west] at (-8.5,-.4) {\large Games under\\\large  delayed control};
        \node[align=left,anchor=west] at (-8.5,-3) {\large Delay games};
        
    \end{tikzpicture}
    \caption{The relation between games under delayed control and delay games with Borel winning conditions. The upper ellipsis contains pairs~$(\arenagame, \delta)$ consisting of a game~$\arenagame$ under delayed control and a fixed delay~$\delta$; the lower one contains delay games~$\delaygame{L}$ for some fixed~$k$. 
    The arrows represent the two transformations described in \autoref{sec:transformation}.}
    \label{fig:transf}
\end{figure}
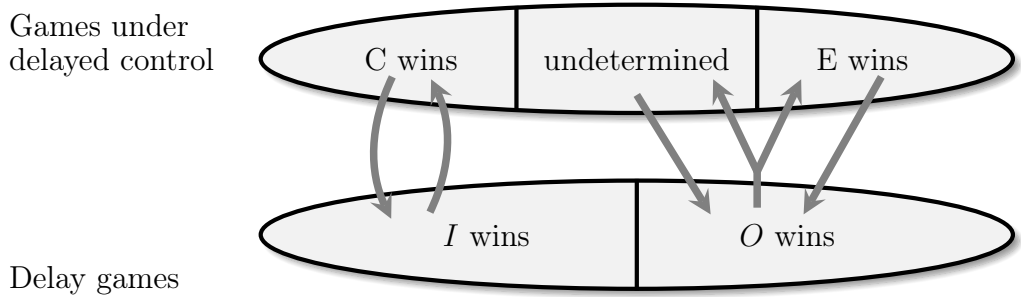

The full relation between games under delayed control and delay games is depicted in \autoref{fig:transf}, restricted to Borel winning conditions (note that both transformations described in \autoref{sec:transformation} preserve Borelness).
The equivalence between controller winning the game under delayed control and Player~$I$ winning the corresponding delay game has been shown in \autoref{lemma:fromDelayedControlToDelay} and \autoref{lemma:fromDelayToDelayedControl}.
Also, \autoref{lemma:fromDelayToDelayedControl} and \autoref{remark:transformationsareinverses} imply that undetermined safety games under delayed control and those won by environment get transformed into delay games that are won by Player~$O$.
\autoref{lemma:envtoinput} shows that games under delayed control won by environment are transformed into delay games won by Player~$O$ while chaining \autoref{lemma:fromDelayedControlToDelay} and \autoref{lemma:fromDelayToDelayedControl} shows that undetermined games under delayed control are also transformed into delay games won by Player~$O$.
Finally, \autoref{lemma:fromDelayedControlToDelay} and \autoref{remark:transformationsareinverses} imply that delay games won by Player~$O$ get transformed into undetermined safety games under delayed control or to ones that are won by environment (and it straightforward to construct delay games that realize both cases).

\section{Refining the Correspondence: Sure Winning and Almost Sure Winning}
\label{Sec:AlmostSure}

It should be noted that the above transformations of games under delayed control into delay games and vice versa hinge on the fact that environment in the game under delayed control could, though lacking recent state information to do so strategically, by mere chance play the very same actions that the informed Player~$O$ in the delay game plays in his optimal adversarial strategy. 
That this constitutes a fundamental difference becomes apparent if we consider almost sure winning instead of sure winning. Almost sure winning calls for the existence of a mixed strategy that wins with probability~$1$, i.e., may fail on a set of plays with measure~$0$. This is different from sure winning in the sense of the definition of winning strategies for games under delayed control in \autoref{subsec:games}, which calls for a strategy that never fails.

\begin{rem}
We introduce mixed strategies for games under delayed control only, as delay games (with Borel winning conditions) are determined, which means that mixed strategies do not offer any advantage over pure strategies as introduced in \autoref{subsec:delaygames}.
\end{rem}

Given an even $\delta \ge 0$, a mixed strategy for controller in $\arenagame$ under delay~$\delta$ is a pair $(\alpha, \stratC)$ where $\alpha \in \prob{(\Sigma_c)^{\frac{\delta}{2}}}$ is a probability distribution over $(\Sigma_c)^{\frac{\delta}{2}}$ and $\stratC \colon \prefsC \rightarrow \prob{\Sigma_c}$ maps play prefixes ending in $S_c$ to probability distributions over actions of controller.
A mixed strategy for environment is a mapping~$\stratE\colon \prefsE \rightarrow \prob{\Sigma_e}$.

The notion of consistency of a play with a strategy simply carries over, now inducing a Markov chain due to the probabilistic nature of the strategies. We say that a mixed strategy for controller (environment) \emph{wins almost surely} if and only if it wins against any strategy of its opponent environment (controller) with probability~$1$, i.e., if and only if the winning condition is satisfied with probability~$1$ over the Markov chain induced by the game and the particular strategy combination. In this section, we write sure winning for winning as defined in \autoref{sec:prels}, as is usual for games with randomized strategies.

The notion of almost sure winning alters chances for the players substantially by excluding the possibility of reliably playing an optimal strategy though lacking the information for doing so due to delayed observation. This can be seen from the following lemma, stating a fundamental difference between controller's power in games under delayed control and Player~$I$'s power in the corresponding delay games.

\begin{lem}\label{lemma:almost-sure-win-vs-sure-loss}
    There is a game~$\arenagame$ under delayed control and some (even) delay~$\delta$ such that controller wins $\arenagame$ under delay~$\delta$ almost surely while Player~$O$ (not Player~$I$, which is the player corresponding to controller) wins the corresponding delay game~$\delaygame{\complement{L(\arenagame)}}$ for $k = \frac{\delta}{2}$, and surely so. 
\end{lem}
 
\begin{proof}
    Consider the reachability game in \autoref{fig:C-almost-surely-vs-O-surely} under delay~$2$ (or any larger delay). 
    Intuitively, the players place a coin in each round (by picking either heads to tails with each move) and controller wins a play if the black state is visited, which happens if she selects a different coin placement than chosen by environment in the previous move.

\begin{figure}
    \centering
\begin{tikzpicture}[ultra thick,yscale=1.3]
    
     \draw[gray!20, rounded corners,line width=2mm] (-1.15,-1.3) rectangle (10.5,1.4);  
    
    \draw[fill=gray!20,gray!20] (1.5,-1.3) rectangle (4.5,1.4); 
    \draw[fill=gray!20,gray!20] (7.5,-1.3) rectangle (10.5,1.4); 
    
    \node[black] at (0,1.1) {\small \bf C};
    \node[black] at (3,1.1) {\small \bf E};
    \node[black] at (6,1.1) {\small \bf C};
    \node[black] at (9,1.1) {\small \bf E};

     \node[mystate] (start) at (-0,0) {\phantom{v}};
     \node[mystate] (et) at (3,0) {\phantom{v}};
     \node[mystate,fill] (eb) at (9,0) {\phantom{v}};
     \node[mystate] (ct) at (6,.5) {\phantom{v}};
     \node[mystate] (cb) at (6,-.5) {\phantom{v}};
    
    \path[-stealth]
    (-.75,0) edge (start)

     (start) edge node[above,near start] {$h,t$} (et)  
     (et.north) edge[bend left=20,near start] node[above] {$h$} (ct)  
     (et.south) edge[bend right=20,near start] node[below] {$t$} (cb)  
     (eb) edge[bend left=0,near start] node[above] {$h$} (ct)  
     (eb) edge[bend left=0,near start] node[below] {$t$} (cb)  

     (ct) edge[bend left=0,very near start] node[below] {$h$} (et)  
     (ct) edge[bend left=20, very near start] node[above] {$t$} (eb.north)  
     (cb) edge[bend right=0, very near start] node[above] {$t$} (et)  
     (cb) edge[bend right=20, very near start] node[below] {$h$} (eb.south);

\end{tikzpicture}
    \caption{A reachability game that, under any positive delay, is won by controller almost surely via the simple randomized strategy of coin tossing (thus randomly generating head and tail events $h$ and $t$), but won by player $O$ surely if interpreted as a delay game due to the lookahead on Player~$I$'s actions granted to Player~$O$. The initial state is marked by an arrow and controller wins if and only if the black vertex is visited at least once.}
    \label{fig:C-almost-surely-vs-O-surely}
\end{figure}
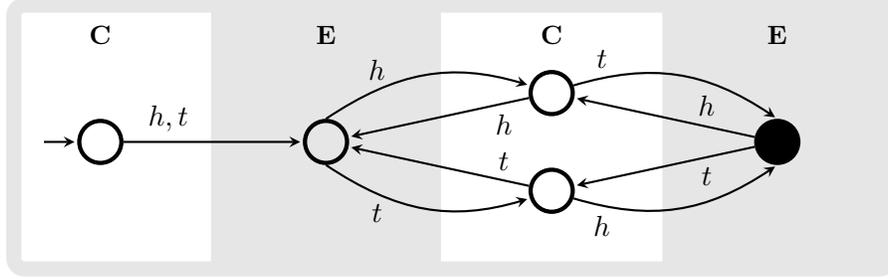

    Under any even (by definition) positive delay, controller wins this game with probability~$1$, i.e., almost surely, by a simple randomized strategy of coin tossing: in each step randomly selecting action~$h$ or $t$ with positive probability each, an eventual visit of the black state is guaranteed with probability~$1$, irrespective of being uninformed about environment's preceding move due to the delay.

    The corresponding delay game~$\delaygame{\complement{L(\arenagame)}}$ for $k = \frac{\delta}{2}$, however, is easily won by Player~$O$, because in delay games, the delayed Player~$I$ grants a lookahead to Player~$O$. Hence, Player~$O$ can, due to the delay, already see the next move of Player~$I$ such that he can simply copy the next coin placement by Player~$I$, safely staying in the non-black states and thereby win. 
\end{proof}
\noindent 
Note that \autoref{lemma:almost-sure-win-vs-sure-loss} implies that the previously observed correspondence between Player~$I$ and controller breaks down when considering almost sure winning strategies instead of just sure winning strategies: 
Games under delayed control for which Player~$O$ wins the corresponding delay game, are no longer either undetermined or won by environment, but may well be won by controller almost surely.

This consequently refines the correspondence between games under delayed control and delay games shown in \autoref{fig:transf} as follows.

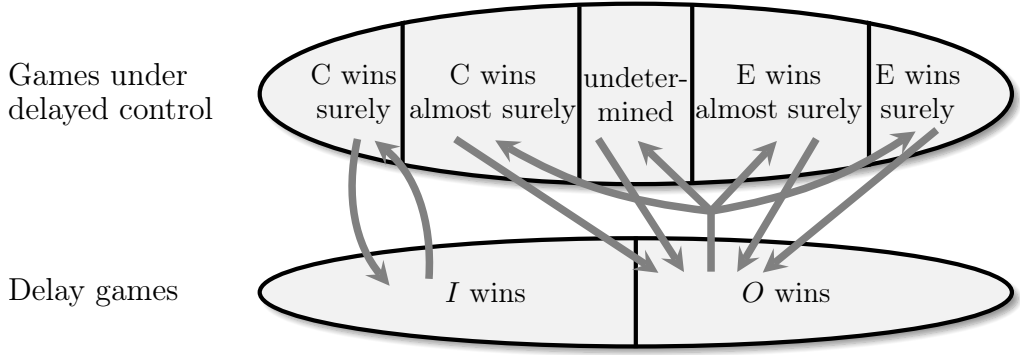
\begin{figure}[t]
    \centering
    \begin{tikzpicture}[ultra thick, yscale=1.2]

        \begin{scope}
            
            \draw[fill=gray!10, blur shadow={shadow blur steps=10}](0,0)circle[x radius=5cm, y radius=1cm];
            \clip(0,0)circle[x radius=5.1cm, y radius=1.019cm];
             \draw[] (-.75,-1) -- (-.75,1);
             \draw[] (.75,1) -- (.75,-1);
             \draw[] (-3.1,-1) -- (-3.1,1);
             \draw[] (3.1,1) -- (3.1,-1);
        \end{scope}

        \begin{scope}
           
            \draw[fill=gray!10, blur shadow={shadow blur steps=10}](0,-2.2)circle[x radius=5cm, y radius=.6cm];
            \clip(0,-2.2)circle[x radius=5.1cm, y radius=.619cm];        
            \draw[] (0,-1) -- (0,-4.5);
        \end{scope}
        
        \node[align = center] at (0,0) {undeter-\\ mined};
        \node[align = center] at (-3.75,0) {C wins\\ surely};
        \node[align = center] at ( 3.75,0) {E wins\\ surely};
        \node[align = center] at (-1.9,0) {C wins\\ almost surely};
        \node[align = center] at ( 1.9,0) {E wins\\ almost surely};
        \node[] at (-2, -2.2) {$I$ wins};
        \node[] at ( 2, -2.2) {$O$ wins};

        \path[-stealth]
        (-3.7,-.5) edge[line width= 1mm,bend right,gray] (-3.25,-2.15)
        (-2.75,-2.05) edge[line width= 1mm,bend right,gray] (-3.5,-.5)
        
        (4,-.4) edge[line width= 1mm,gray] (1.66,-2)
        (2.4,-.5) edge[line width= 1mm,gray] (1.33,-2)
        (-0.5,-.5) edge[gray, line width=1mm] (.66,-2)
        (-2.4,-.5) edge[gray, line width=1mm] (.33,-2)

        (1,-1.3) edge[line width= 1mm,gray,-] (1,-2)
        (1,-1.3) edge[bend left = 10, line width= 1mm,gray] (-1.9,-.5)
        (1,-1.3) edge[line width= 1mm,gray] (0,-.5)
        (1,-1.3) edge[line width= 1mm,gray] (1.9,-.5)
        (1,-1.3) edge[bend right = 10, line width= 1mm,gray] (3.75,-.4)
        ;
        
        \node[align=left,anchor=west] at (-8.5,0) {\large Games under\\\large delayed control};
        \node[align=left,anchor=west] at (-8.5,-2.2) {\large Delay games};
        
    \end{tikzpicture}
    \caption{The relation between safety games under delayed control and delay games with Borel winning conditions. The upper ellipsis contains pairs~$(\arenagame, \delta)$ consisting of a game~$\arenagame$ under delayed control and a fixed delay~$\delta$; the lower one contains delay games~$\delaygame{L}$ for some fixed~$k$. 
    The arrows represent the two transformations described in \autoref{sec:transformation}.}
    \label{fig:refined-correspondence}
\end{figure}

\begin{thm}
\label{theorem:refined-correspondence}
    Given a game~$\arenagame$ with Borel winning condition and an even $\delta \ge 2$,  the following correspondences between $\arenagame$ and the corresponding delay game~$\delaygame{\complement{L(\arenagame)}}$ for $k = \frac{\delta}{2}$ hold:
    \begin{enumerate}
        \item\label{item:csure} Controller surely wins $\arenagame$ under delay~$\delta$ if and only if Player~$I$ surely wins $\delaygame{\complement{L(\arenagame)}}$.

        \item\label{item:calmostsure} If controller almost surely wins $\arenagame$ under delay $\delta$ but cannot surely win $\arenagame$ under delay $\delta$ then Player~$O$ surely wins $\delaygame{\complement{L(\arenagame)}}$.
        
        \item\label{item:e} If environment surely or almost surely wins $\arenagame$ under delay $\delta$ then Player~$O$ wins $\delaygame{\complement{L(\arenagame)}}$.
        
        \item\label{item:undet} If $\arenagame$ is undetermined under delay $\delta$ with respect to  almost sure winning strategies then Player~$O$ wins $\delaygame{\complement{L(\arenagame)}}$.
        
        \item\label{item:nonempty} All the aforementioned classes are non-empty, i.e., there exist games under delayed control where controller wins, where controller wins almost surely (but not surely), where environment wins surely, where environment wins almost surely (but not surely), and games which are undetermined with respect to  almost-sure winning strategies.
    \end{enumerate}
    The above correspondences are depicted in \autoref{fig:refined-correspondence}.
\end{thm}

\begin{proof}
    Equivalence~\eqref{item:csure} has been shown in \autoref{sec:results}. Implications~\eqref{item:calmostsure}--\eqref{item:undet} follow immediately from \eqref{item:csure} and determinacy of the delay game~$\delaygame{\complement{L(\arenagame)}}$: whenever controller fails to have a sure winning strategy then Player~$I$ does not have one either, and Player~$O$ consequently wins surely due to determinacy of $\delaygame{\complement{L(\arenagame)}}$.

    For \eqref{item:nonempty}, one observes that games where either controller or environment surely win exist trivially, as universal as well as empty winning conditions are permitted. Existence of a game under delayed control that can be won by controller almost surely, but not surely, is shown by the example in \autoref{fig:C-almost-surely-vs-O-surely}. 
    Complementation of its winning condition, i.e., declaring environment the winner if and only if the black vertex is visited, converts this game into one that environment wins almost surely, but not surely. Existence of a game that is undetermined with respect to  almost sure winning strategies is witnessed by the game depicted in \autoref{fig:env}, wherein under delay~$\ge 2$, neither controller nor environment can secure a win with probability strictly larger than $\frac{ 1}{ 2}$. 
    This game consequently remains undetermined with respect to  almost sure winning strategies.
\end{proof}
\noindent 
Item~\eqref{item:calmostsure} of \autoref{theorem:refined-correspondence} is of particular interest, as it expresses a delay-related strengthening of controller relative to Player~$I$, letting controller win almost surely where Player~$I$ looses for sure.
The correspondence between controller and Player~$I$ observed in the deterministic setting thus breaks down when almost sure winning is considered and mixed strategies are permitted.

\section{Further Refining the Correspondence: Winning with Probability}
\label{ Sec:Probability}

The differences between delay games and games under delayed control become even more pronounced when we ask for strategies that win with at least a given probability $\theta \in ]0,1[$ instead of asking for an almost-sure win.

\begin{figure}
    \centering
\begin{tikzpicture}[ultra thick,yscale=1.3, xscale=.57]

    \draw[gray!20, rounded corners,line width=2mm] (-1.15,-2.4) rectangle (25.15,1.7);  
    
    \draw[fill=gray!20,gray!20] (1.5,-2.3) rectangle (4.5,1.7); 
    \draw[fill=gray!20,gray!20] (7.5,-2.3) rectangle (10.5,1.7); 
    \draw[fill=gray!20,gray!20] (13.5,-2.3) rectangle (16.5,1.7); 
    \draw[fill=gray!20,gray!20] (19.5,-2.3) rectangle (22.5,1.7); 
    
    \node[black] at (0,1.4) {\small \bf C};
    \node[black] at (3,1.4) {\small \bf E};
    \node[black] at (6,1.4) {\small \bf C};
    \node[black] at (9,1.4) {\small \bf E};
    \node[black] at (12,1.4) {\small \bf C};
    \node[black] at (15,1.4) {\small \bf E};
    \node[black] at (18,1.4) {\small \bf C};
    \node[black] at (21,1.4) {\small \bf E};
    \node[black] at (24,1.4) {\small \bf C};

    \node[mystate] (1) at (0,0.5) {\phantom{v}};

    \node[mystate] (2) at (3,0.5) {\phantom{v}};
    \node[mystate] (3) at (6, .5) {\phantom{v}};
    \node[mystate] (4) at (6,-.5) {\phantom{v}};
    \node[mystate,fill = black] (5) at (3,-1.5) {\phantom{v}};
    \node[mystate,fill = black] (6) at (0,-1.5) {\phantom{v}};

    \node[mystate] (7) at (9, .5) {\phantom{v}};
    \node[mystate] (8) at (12, .5) {\phantom{v}};
    \node[mystate] (9) at (12, -.5) {\phantom{v}};
    \node[mystate,fill = black] (10) at (9,-1.5) {\phantom{v}};
    \node[mystate,fill = black] (11) at (6,-1.5) {\phantom{v}};

    \node[mystate] (12) at (15, .5) {\phantom{v}};
    \node[mystate] (13) at (18, .5) {\phantom{v}};
    \node[mystate] (14) at (18, -.5) {\phantom{v}};
    \node[mystate,fill = black] (15) at (15,-1.5) {\phantom{v}};
    \node[mystate,fill = black] (16) at (12,-1.5) {\phantom{v}};

    \node[mystate] (17) at (21, .5) {\phantom{v}};
    \node[mystate] (18) at (24, .5) {\phantom{v}};

    \path[-stealth]

    (-1,0.5) edge (1)
    (1) edge[near start] node[above] {$h,t$} (2)

    (2) edge[near start] node[below] {$t$} (4)
    (2) edge[near start] node[above] {$h$} (3)
    
    (3) edge[near start] node[above] {$h$} (7)  
    (3) edge[ near start] node[above] {$t$} (5.north)  

    (4) edge[near start] node[above] {$t$} (7)  
    (4) edge[near start] node[above] {$h$} (5.north east)  

    (5) edge[bend left=20, near start] node[below] {$h,t$} (6)  
    (6) edge[bend left=20, near start] node[above] {$h,t$} (5)  

    (7) edge[near start] node[below] {$t$} (9)
    (7) edge[near start] node[above] {$h$} (8)
    
    (8) edge[near start] node[above] {$h$} (12)  
    (8) edge[ near start] node[above] {$t$} (10.north)  

    (9) edge[near start] node[above] {$t$} (12)  
    (9) edge[near start] node[above] {$h$} (10.north east)  

    (10) edge[bend left=20, near start] node[below] {$h,t$} (11)  
    (11) edge[bend left=20, near start] node[above] {$h,t$} (10)

    (12) edge[near start] node[below] {$t$} (14)
    (12) edge[near start] node[above] {$h$} (13)
    
    (13) edge[near start] node[above] {$h$} (17)  
    (13) edge[ near start] node[above] {$t$} (15.north)  

    (14) edge[near start] node[above] {$t$} (17)  
    (14) edge[near start] node[above] {$h$} (15.north east)  

    (15) edge[bend left=20, near start] node[below] {$h,t$} (16)  
    (16) edge[bend left=20, near start] node[above] {$h,t$} (15)  

    (17) edge[bend left=20, near start] node[above] {$h,t$} (18)  
    (18) edge[bend left=20, near start] node[below] {$h,t$} (17)

    ;

\end{tikzpicture}
    \caption{A reachability game that, under any positive delay, is won by controller with probability $\frac 7 8$ via the simple randomized strategy of balanced coin tossing (thus randomly generating head and tail events $h$ and $t$ with probability $\frac 1 2$ each), but won by player O surely if interpreted as a delay game due to the lookahead on Player~$I$’s actions granted to Player~$O$. The objective for controller (Player~$I$, resp.) in this  game is to reach some black state.}
    \label{fig:Probability}
\end{figure}
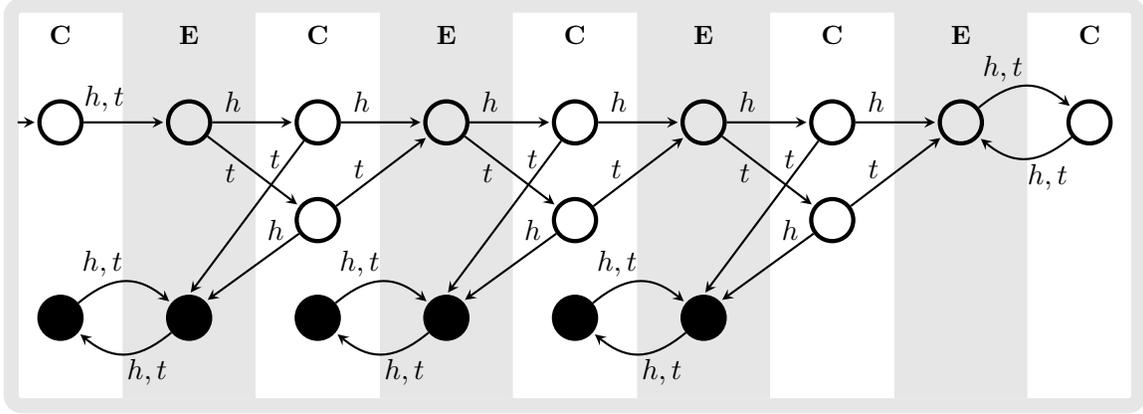
\autoref{fig:Probability} depicts a reachability game under delayed control that controller can win with probability~$\frac 1 2 + \left(\frac 1 2\right)^2 + \left(\frac 1 2\right)^3 = \frac 7 8$ when the delay is non-zero, while Player~$I$ in the corresponding delay game looses surely.

Such a difference is not particular to reachability games: Note that controller would also win the Büchi game defined by the black states being controller's winning set with probability $\frac 7 8$, and likewise controller would win the safety game where it has to avoid the black states forever with probability $\frac 1 8$, while all these games will surely be lost by Player~$I$.
Note that we can furthermore also generate a safety game that is won by controller with probability $\frac 7 8$ (and likewise a reachability game or a Büchi game which are won by controller with probability $\frac 1 8$) by simply swapping the acceptance status of the four bottom connected components in the game graph of \autoref{fig:Probability}. These games are still lost by Player~$I$ for sure.

Our next result shows that a game with these properties exists for every rational number in $[0,1]$.

\begin{lem}\label{Lemma:rational-probabilities}
    Let $\theta \in [0,1]$ be rational. Then there exists a game under delayed control with $\omega$-regular winning condition that controller wins with probability~$\theta$.
\end{lem}
\begin{proof}
    The case of $\theta = 1$ has been covered in  \autoref{sec:transformation} and \autoref{Sec:AlmostSure}, as all games won by Player~$I$ as well as the game in \autoref{fig:C-almost-surely-vs-O-surely} are won with probability 1 by controller. Hence, w.l.o.g., $\theta = \frac n m$ with $n, m$ being integer and $0 \le n < m$. Consider the game where first controller picks a number $a \in \{0,\ldots,m-1\}$ and then environment picks a number $b \in \{0,\ldots,m-1\}$. The game then is over (i.e., the states now reached are all sinks) and controller  wins if and only if $(a+b) \mod m < n$. 
    
    When not subject to delay, this game is won by the environment by observing $a$ and picking $b = m-1-a$, which gives $(a+b) \mod m = (a + m-1-a) \mod m = m-1 \ge n$, violating controller's winning condition. 
    
    But under any positive delay, the best strategy played by controller obviously is to draw a random choice uniformly from $\{0,\ldots,m-1\}$, as any biased choice (no matter whether deterministic or random) would render its otherwise unobservable (due to the delay) choice predictable to the environment and thus give the environment an advantage. But with $a$ being chosen uniformly from $\{0,\ldots,m-1\}$, the random variable $(a+b) \mod m$ also is uniformly distributed over $\{0,\ldots,m-1\}$, no matter what strategy the environment plays, as the environmental strategy has to be independent of the particular choice of $a$ due to the delay. The chance for controller to win, given its winning condition $(a+b) \mod m < n$ and the uniform distribution of $(a+b) \mod m$ over $\{0,\ldots,m-1\}$, consequently is $\frac n m$. Hence, this game under delayed control features a win probability of exactly $\frac n m = \theta$ for controller.
\end{proof}

\begin{rem}
    By the symmetry between controller and environment inherent to games under delayed control with respect to a fixed delay~\cite[Section 2]{ChenFLMZ21}, which subjects both to the same informedness constraints, environment wins with probability~$1-\theta$ when controller wins with probability~$\theta$.
Hence, for each $\theta \in [0,1]$ being rational there also exists a game under delayed control with $\omega$-regular winning condition that environment wins with probability $\theta$. 
\end{rem}

As delay games with Borel winning conditions are determined, there obviously is no counterpart to the property expressed in \autoref{Lemma:rational-probabilities} for delay games: in such delay games, the winning probabilities are, under optimal strategies, always either 0 or 1; in fact, these losses or wins are even sure and not only almost sure.
\begin{lem}\label{lem:less-than-one}
    If a game under delayed control with Borel winning condition is won by controller with probability $\theta < 1$ then the corresponding delay game is won by Player~$O$ surely.
\end{lem}
\begin{proof}
    As $\theta < 1$, the controller cannot win the game surely. By the results from \autoref{sec:transformation}, in particular \autoref{lemma:fromDelayToDelayedControl}, this implies that Player~$I$ cannot win the corresponding delay game. As delay games with Borel winning conditions are determined, Player~$O$ then wins (for sure).
\end{proof}
\noindent 
Please note that in case of a win probability of $\theta = 1$ for controller, either a win of Player~$I$ or of Player~$O$ may apply, as has been shown in \autoref{sec:transformation} and \autoref{Sec:AlmostSure}. The transformation from \autoref{sec:transformation} shows that games won by Player~$I$ are surely won by controller, while \autoref{fig:C-almost-surely-vs-O-surely} provides an example of a game won surely by Player~$O$ while won with probability 1 by controller.

\begin{figure}
    \begin{tikzpicture}[thick]

 \clip(-6.217,-1.5) rectangle (6.217,1.5);
\begin{scope}
\draw[fill=gray!10, blur shadow={shadow blur steps=10}](0,0)circle[x radius=15cm, y radius=1.3cm];

\node[gray!70,rotate=20] at (0,0) {\Huge undetermined};

    \clip[](0,0)circle[x radius=15cm, y radius=1.3cm];

   \node at (0,0) {$\frac 1 2$};
\node at (3.2,0) {$\frac 3 4$};
\node at (1.6,0) {$\frac 5 8$};
\node at (4.8,0) {$\frac 7 8$};
\node at (-3.2,0) {$\frac 1 4$};
\node at (-1.6,0) {$\frac 3 8$};
\node at (-4.8,0) {$\frac 1 8$};
\node at (-5.6,0) {$\frac{1}{16}$};
\node at (-4,0) {$\frac{3}{16}$};
\node at (-2.4,0) {$\frac{5}{16}$};    
\node at (-.8,0) {$\frac{7} {16}$};    
\node at (5.6,0) {$\frac{15}{16}$};
\node at (4,0) {$\frac{13}{16}$};
\node at (2.4,0) {$\frac{11}{16}$};    
\node at (.8,0) {$\frac{9}{16}$};

    \foreach \x in {.2, .6, ..., 6.4}{
        \draw(\x,1.4) -- (\x,-1.4);
}
    \foreach \x in {-.2, -.6, ..., -6.4}{
        \draw(\x,1.4) -- (\x,-1.4);
}


\foreach \s in {-6.4,-5.6,...,6.4}{
    
        \node at (\s-.38,0) {\scriptsize$\cdots$};

}

\end{scope}
    
    \end{tikzpicture}
    \caption{A zoom on the area labeled \myquot{undetermined} in \autoref{fig:refined-correspondence}. For every rational number~$\theta \in ]0,1[$ there is a game under delayed control~$\arenagame$ and a delay~$\delta$ such that controller wins $\arenagame$ under delay~$\delta$ with value~$\theta$.}
    \label{fig:zoomed-correspondence}
\end{figure}
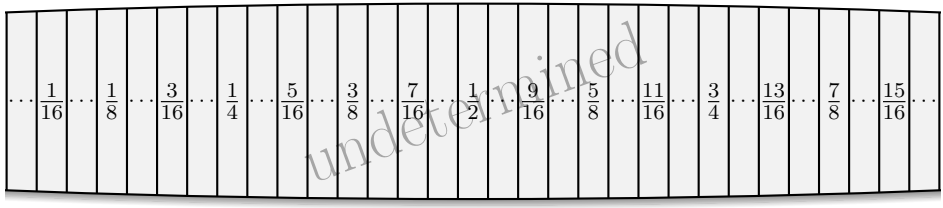

As \autoref{Lemma:rational-probabilities} demonstrates that the win probabilities of games under delayed control are dense in $[0,1]$, the following refined statement about correspondence between the different types of games under delay makes sense.
\begin{thm}\label{thm:Non-zreo-one-probabilities}
    Let $\arenagame$ be a game with Borel winning condition, $\delta = 2k$ with $k \in \nats\setminus\{0\}$ a delay, and $\theta \in [0,1]$ the probability of winning $\arenagame$ under delay~$\delta$ by  controller with an optimal mixed strategy. 
    \begin{enumerate}
    
        \item\label{item:non-one-probability} If $\theta < 1$ then Player~$O$ wins the corresponding delay game $\delaygame{\complement{L(\arenagame)}}$ with delay $k$, and this win is for sure.
    
        \item\label{item:probability-one} If $\theta = 1$, yet controller does not win $\arenagame$ surely, then Player~$O$ wins in the corresponding delay game $\delaygame{\complement{L(\arenagame)}}$ with delay $k$ for sure.
        
        \item\label{item:sure-win} If controller wins $\arenagame$ surely under delay $\delta$ then Player~$I$ wins the corresponding delay game $\delaygame{\complement{L(\arenagame)}}$ for sure under delay $k$.
        
        \item\label{item:sure-loss} If environment wins $\arenagame$ surely under delay $\delta$ then Player~$O$ wins the corresponding delay game $\delaygame{\complement{L(\arenagame)}}$ for sure under delay $k$.
        
        \item\label{item:nonempty-prob} All the aforementioned classes are non-empty, i.e., there exist games under delayed control where controller wins, where controller wins almost surely (but not surely), where controller wins with some probability $\theta \in\, ]0,1[\,$, where environment wins almost surely  (but not surely), and where environment wins surely.
    \end{enumerate}
\autoref{fig:zoomed-correspondence} refines the are labeled  \myquot{undetermined} in \autoref{fig:refined-correspondence}.
\end{thm}
\begin{proof}
    \autoref{lem:less-than-one} states property \eqref{item:non-one-probability}. Properties  \eqref{item:probability-one} to \eqref{item:sure-loss} are reformulations of the corresponding properties from \autoref{theorem:refined-correspondence}. The last property \eqref{item:nonempty-prob} follows from the corresponding property in \autoref{theorem:refined-correspondence} together with \autoref{Lemma:rational-probabilities}.
\end{proof}

\section{Conclusion}
\label{sec:conc}

We have compared delay games~\cite{KleinZ14} and games under delayed control~\cite{ChenFLMZ21}, two types of infinite games aiming to model asynchronicity in reactive synthesis,  and have exhibited the differences in definitions and charted the relation between them with respect to  both deterministic and randomized strategies: 
When considering sure winning and deterministic strategies, one can efficiently transform a game under delayed control into a delay game such that controller wins the game under delayed control with delay~$\delta$ by a deterministic strategy if and only if Player~$I$ wins the resulting delay game with lookahead of size~$\frac{\delta}{2}$. 
Dually, one can efficiently transform a delay game into a game under delayed control such that Player~$I$ wins the delay game with lookahead of size~$k$ if and only if controller wins the resulting game under delayed control with delay~$2k$ by a deterministic strategy. 
These results allow us to transfer known complexity results and bounds on the amount of delay from delay games to games under delayed control, for which no such results were known (when considering deterministic strategies).

Analogous results fail in the setting of randomized strategies and almost sure or even quantitative winning conditions. Here, it is completely open who can win the game under delayed control and with which probability, unless Player~$I$ can win the delay game surely. For the case that Player~$I$ does not have a sure winning strategy, i.e., whenever Player~$O$ has a sure winning strategy (due to determinacy of delay games with Borel winning conditions), we can construct games that controller, normally corresponding to Player~$I$, nevertheless wins almost surely, wins with any rational probability in $]0,1]$, loses almost surely, or loses surely. These findings refine our original result from \cite{ChenFLMZ21} that games under delayed control are not determined. They also expose a profound difference between the problem of control via delaying channels, which games under delayed control  capture, and delay games.

\paragraph*{Acknowledgements:}
Martin Fränzle has been supported by Deutsche Forschungsgemeinschaft under grant no.\ DFG FR 2715/5-1 ``Konfliktresolution und kausale Inferenz mittels integrierter sozio-technischer Modellbildung''. Paul Kröger received funding from Germany's Federal Ministry of Education and Research strategic ``MANNHEIM'' project ``\mbox{AutoDevSafeOps}---Integrated development and operation of safe automotive systems''.
The majority of the research was carried out while Sarah Winter has been affiliated with Université libre de Bruxelles, Belgium and has been supported by the Fonds National de la Recherche Scientifique – F.R.S.-FNRS.
Martin Zimmermann has been supported by DIREC---Digital Research Centre Denmark.

\bibliographystyle{alphaurl}
\bibliography{bib}

\end{document}